\documentclass[useAMS,usenatbib]{mn2e}
\usepackage{psfig}
\usepackage{amsmath}

\def \bh {_{\rm BH}}
\def \mbh {M_{\rm BH}}
\def \blr {_{\rm BLR}}

\def \mgii {Mg\,{\sc ii}}
\def \civ {C\,{\sc iv}}
\def \hbeta {H$\beta$}

\def \pmo {$^{-1}$}

\def \empha {({\emph a})}
\def \emphb {({\emph b})}
\def \emphc {({\emph c})}
\def \emphd {({\emph d})}
\def \sn {{\rm S/N}}

\def \msun {M$_{\odot}$}

\def \hb {H$\beta$}
\def \bj {b_{\rm J}}

\def \civ {C\,{\sc iv}}
\def \heii {He\,{\sc ii}}
\def \oiii {O\,{\sc iii]}}
\def \niv {N\,{\sc iv]}}
\def \siii {Si\,{\sc ii}}
\def \feii {Fe\,{\sc ii}}

\def \empha {({\emph a})}
\def \emphb {({\emph b})}
\def \emphc {({\emph c})}
\def \emphd {({\emph d})}

\def \mgiie {Mg\, {\scriptscriptstyle II}}
\def \cive {C\, {\scriptscriptstyle IV}}

\title[The {\rm \civ} line width distribution]{The \civ\ line width
  distribution for quasars and its implications for broad-line region
  dynamics and virial mass estimation.}

\author[S. Fine et al.]
       {S. Fine$^{1,2}$\thanks{stephen.fine@durham.ac.uk},
         S.~M. Croom$^1$, J. Bland-Hawthorn$^1$, K.~A. Pimbblet$^3$,
       N.~P. Ross$^{2,4}$,
	 \newauthor D.~P. Schneider$^{4}$, T. Shanks$^{2}$ \\
$^1$Sydney Institute for Astronomy, School of Physics, The University
       of Sydney, NSW 2006, Australia. \\
$^{2}$Department of Physics, Durham University, South Road, Durham DH1
      3LE, UK.\\
$^3$School of Physics, Monash University, Clayton, Victoria 3800, Australia \\
$^4$Department of Astronomy and Astrophysics, The Pennsylvania State
      University, 525 Davey Laboratory, University Park, PA 16802,
       USA.\\
}

\begin{document}

\maketitle

\begin{abstract}

We perform an extensive analysis of the \civ\,$\lambda1549$ line in
three large spectroscopic surveys of quasars. Differing
approaches for fitting the \civ\ line can be found in the literature, and
we compare the most common methods to highlight the relative
systematics associated with each. We choose the line fitting
procedure that results in a symmetric profile for the \civ\ line and
gives accurate fits to local emission features around the line,
and use this approach to measure the width of the \civ\ line
in spectra from the SDSS, 2QZ and 2SLAQ surveys.

The results are compared with a previous study of
the \mgii\,$\lambda2799$ line in the same sample. We find the \civ\
line tends to be broader than the \mgii\ line in spectra that have both
lines, and the average ratio between the lines is consistent with a 
simplistic model for a photoionised, virialised and stratified
broad-line region. There exists a statistically significant correlation between
the widths of the \civ\ and \mgii\ lines. However, the correlation is weak,
and the scatter around a best fit is only marginally less than
the full dynamic range of line widths.

Motivated by previous work on the \mgii\ line, we examine the
dispersion in the distribution of \civ\ line widths. We find that the
dispersion in \civ\ line widths is essentially independent of both
redshift and luminosity. This result is in stark contrast to the
\mgii\ line, which shows a strong luminosity dependence. Furthermore we
demonstrate that the low level of dispersion in \civ\ line width
($\sim0.08$\,dex) is inconsistent with a pure-disk model for the
emitting region and use our data to constrain simple models for the
broad-line region.

Finally we consider our results in terms of their implications for the
the virial technique for estimating black hole masses. The
inconsistency between \mgii\ and \civ\ line widths in single spectra,
combined with the differing behaviour of the \mgii\ and \civ\ line
width distributions as a whole, indicates that there must be an
inconsistency between \mgii\ and \civ\ virial mass
estimators. Furthermore, the level of intrinsic
dispersion in \mgii\ and \civ\ line widths contributes less
dynamic range to virial mass estimates than the error associated with
the estimates. The indication is that the line width
term in these UV virial mass estimators may be essentially irrelevant
with respect to the typical uncertainty on a mass estimate.

\end{abstract}


\section{Introduction}

This paper describes an extensive analysis of the \civ\,$\lambda1549$
line width distribution in three large spectroscopic samples of
quasi-stellar objects (QSOs).
\civ\ has the highest ionisation potential of any of the strong broad
emission lines in QSO spectra and is the best probe of the high-ionisation
inner regions of the broad-line region (BLR). Furthermore, like \hb\
and \mgii\,$\lambda2799$, the \civ\ line is commonly used to calculate
super-massive black hole (SMBH) masses for QSOs. The line is
observed in optical spectra with redshifts between $\sim$1.5 and 5
(corresponding to a look-back time of 9 to 12\,Gyr), and is the only line
used to calculate SMBH masses in the highest redshift objects.

The redshift range between $\sim$1.5 and 5 is of particular importance to
quasar astrophysics since 
it spans the so called `quasar epoch' at $z\sim2$ to $3$. Before $z\sim3$ the
space density of QSOs has been observed to increase with time
(e.g. \citealt{osm82,fan01,rich06b}), but since $z\sim2$ it has fallen
(e.g. \citealt{lon66,sch68,croom04,rich06b}). The period between $z\sim2$ and
$3$ marks the peak of quasar activity in the Universe. Understanding the
processes which caused this ramping up of activity in the early
Universe and what is responsible for its reversal is a major
goal in QSO science. The fact that the \civ\ line
is visible in optical spectra over this entire range potentially makes it an
attractive probe of QSOs at these epochs.
%
%
%
%

In section~\ref{sec:vir_mass} we review virial SMBH mass estimation
and the use of the \civ\ line with this technique.
Section~\ref{sec:data} gives a brief description of the three
datasets used in this work.
In section~\ref{sec:results} we present the results of our fitting and
make a comparison with a previous analysis of the \mgii\ line in the
same sample of quasars from \citet{me2}. Motivated by the results in
\citet{me2}, section~\ref{sec:disp} describes an investigation of the
dispersion in \civ\ 
line widths as a function of redshift and luminosity. In
sections~\ref{sec:vir_wrap} and~\ref{sec:geometry} we discuss our
results with respect to virial SMBH mass estimation and BLR geometry.
Details of the analytic procedure, including a discussion of our line
fitting analysis and a prescription for removing broad-absorption
lines (BALs) from our data, can be found in
appendices~\ref{sec:fitting} and~\ref{sec:bal} respectively.
Throughout this
paper we assume a flat $(\Omega_{\rm m},\Omega_{\Lambda})=(0.3,0.7)$,
$H_{0}=70\,{\rm km\,s}^{-1}\,{\rm Mpc}^{-1}$ cosmology.

\section{\civ\ and virial SMBH mass estimation} \label{sec:vir_mass}

The most common approach for measuring the
mass of QSO SMBHs is through studies of the
BLR. Reverberation mapping \citep{b+m82} allows the size of the BLR to
be derived through studying the time lag between continuum and
broad-line variability in QSO spectra.
Combining an estimate for the size of the BLR with an assumption that
the BLR is virialised,
the mass of the central SMBH can be estimated (e.g. \citealt{pea04}).
Given a QSO with a BLR of radius $r\blr$ and virial velocity $V\blr$
(estimated from the width of an emission line), the central mass is
given by
\begin{equation}
M\bh=f\frac{r\blr V^2\blr}{G}.
\label{equ_virial_real}
\end{equation}
Here the factor $f$ is defined by the geometry and
orientation of the BLR which are unknown (see
e.g. \citealt{p+w99,m+d01,col06,lab06} for discussions on the value of $f$).

Reverberation mapping requires observations over an extended period of time
and as a consequence only a few tens of systems have been adequately
studied in this fashion \citep{kasp00,pea04}. However, in recent years
a technique for estimating SMBH masses from single epoch spectra has
been developed: the `virial' method.

The virial technique for estimating SMBH masses is based on the
radius-luminosity relation measured for the \hbeta\ BLR in
reverberation mapped systems
\citep{wpm99}. This tight correlation between the
continuum luminosity of Seyfert~1s and the \hbeta\ BLR size allows for
single epoch empirical estimation of the BLR size.
The estimated radius is combined with the velocity width of the
\hbeta\ line to give a virial SMBH mass estimate.

The virial mass is estimated with a relation of the form
\begin{equation} \label{equ_virial}
M\bh=A(\lambda L_{\lambda})^{\alpha}FWHM^{2}
\end{equation}
where $FWHM$ is the full width at half maximum of the \hbeta\ line
and $L_{\lambda}$ is the monochromatic luminosity of
the continuum at wavelength $\lambda$ (taken near the line), $A$ is a
normalisation constant, and the
exponent $\alpha$ gives the luminosity dependence of the
radius-luminosity relation.

While the radius-luminosity relation is only well established for the
\hbeta\ BLR, there is growing evidence for a similar relation for
\civ\ \citep{kas07}. Assuming the existence of equivalent
radius-luminosity relations secondary virial mass estimators
based on other emission
lines have been calibrated. Most commonly the \mgii\ or \civ\
lines are used as they are strong, relatively unblended features and
are evident in optical spectra of progressively higher redshift
objects. The relations for these lines also take the form of
equation~\ref{equ_virial} where the FWHM is measured from the new
line, and the continuum luminosity is taken in the vicinity of that
new line. The quantities $A$ and $\alpha$ for these secondary virial estimators
are generally calibrated against SMBH masses measured for the same
sources from the \hbeta\ line. These secondary virial calibrations
have been shown to be consistent with \hbeta\ virial and reverberation
mapping masses to within $\sim0.3$\,dex over several orders of
magnitude in SMBH mass (e.g. \citealt{m+j02,vest02}).

This paper is primarily concerned with the \civ\ line and, 
while the \civ\ line is frequently used to
calculate SMBH masses for QSOs \citep{vest02,v+p06}, there has been
some controversy in
the literature as to how well suited \civ\ is for this sort of
analysis. Most of the debate has focused on the fact that \civ\ is a
considerably higher ionisation line than \hb\ or \mgii, and hence could
be emitted from a different part of the BLR (e.g. \citealt{o+p02}). In
addition,
the \civ\ line profile is known to display asymmetries, due primarily to
absorption in both wings of the line. The \civ\ line also tends to be
blueshifted with respect to lower ionisation lines and narrow lines in QSO
spectra \citep{gas82,rich02} indicating potentially differing dynamics
for the low and high-ionisation BLR. Finally \citet{shen08} found
that, in spectra with both a \mgii\ and \civ\ line, there is little-to-no
correlation between the width of the two lines. \citet{shen08} went on
to conclude that, while virial mass estimators based on the \mgii\ and
\civ\ lines can be inconsistent in individual objects, results averaged
over a population are consistent.

\civ\ virial SMBH mass estimates have been shown
to be consistent with those from other lines as well as for reverberation
mapped objects \citep{v+p06}. In this paper we will
generally assume that \civ\ can be used as a virial mass estimator and derive
our results accordingly (see e.g. \citealt{v+p06,gav08} and
\citealt{b+l05,net07,sul07} for discussions for and against against
this assumption). In section~\ref{sec:vir_wrap} we discuss the
implications of our own findings with respect to the virial assumption
and SMBH mass estimation.

\section{Data and Analysis} \label{sec:data}

We take the same sample studied in \citet{me2}. This comprises of
all of the quasar spectra from the Sloan Digital
Sky Survey (SDSS; \citealt{yor00}) data release five (DR5;
\citealt{ald07}) (as compiled by \citealt{sch07}), the 2dF QSO
Redshift survey (2QZ; \citealt{croom04}) and 2dF SDSS LRG And QSO
survey (2SLAQ; \citealt{ric05,croom09}).  Table~\ref{tab_surveys} shows a 
brief summary of the number of objects and magnitude limits in each
sample. As one moves down the table each successive survey has fewer
spectra, but fainter flux limits. Increasing our flux coverage allows
for a more detailed study of any luminosity effects on the \civ\ line
width distribution.

\begin{table*}
\begin{center}
\caption{Summary of the surveys from which we obtained
  spectra. Successive surveys have fewer spectra but go deeper,
  increasing our luminosity range at a given redshift. Note that the
  magnitude limits quoted for the SDSS QSO survey are those for the
  primary QSO survey. The high redshift sample goes deeper, and
  included are sources observed under different selection criteria and
  also QSOs identified as part of other surveys.}
\label{tab_surveys}
\begin{tabular}{cccccc}
\hline \hline
Survey      & No. of Objects & Mag. Limits & Resolution & Dispersion &
  $\overline{\rm S/N}$ \\
\hline
SDSS (DR5)  & 77,429 & $19>i>15$               & $\sim165$\,km/s &
  $\sim1.5$\,\AA/pix & $\sim13$/pix  \\
2QZ         & 23,338 & $20.85>b_{\rm J}>18.25$ & $\sim465$\,km/s &
  $\sim4.3$\,\AA/pix & $\sim5.5$/pix \\
2SLAQ       &  8,492 & $21.85>g>18.00$         & $\sim465$\,km/s &
  $\sim4.3$\,\AA/pix & $\sim5.5$/pix \\
\hline \hline
\end{tabular}
\end{center}
\end{table*}

\subsection{SDSS spectra}

Details of the SDSS telescope and spectrograph are given in
\citet{gun06} and \citet{sto02}.
The spectra have a logarithmic wavelength scale
translating to a dispersion of
$\sim1-2$\,\AA/pix and a
resolution $\lambda/\Delta\lambda\sim1800$ in the wavelength
range $3800-9200$\,\AA. Objects are observed initially for
2700\,sec. Then are reobserved in 900\,sec blocks until the median
S/N is greater than $\sim4$\,pix\pmo\ resulting in a S/N distribution
with a mean at $\sim13$\,pix\pmo.

The spectra are extracted and reduced with the {\sc spectro2d}
pipeline and automatically classified with {\sc spectro1d}
\citep{sto02}. However,
in creating the Sloan QSO sample used in this paper, \citet{sch07}
visually inspect all of the candidate spectra to determine their
classification.


\subsection{2dF spectra}

Both 2QZ and 2SLAQ spectra were taken with the 2 degree Field (2dF)
instrument on the Anglo-Australian Telescope with the 300B grating
\citep{lewis02}. Spectra have a
dispersion of 4.3\,\AA/pix\ and a resolution of $\sim9$\,\AA\ in
the wavelength range $3700-7900$\,\AA. 2QZ
exposure times were between 3300 and 3600\,sec compared with 14400\,sec
for 2SLAQ. The increase in exposure time for the fainter 2SLAQ
sample results in S/N distributions that are almost
indistinguishable (Both peak at $\sim5.5$\,pix\pmo\ for positive QSO
IDs). 2dF spectra are extracted and manually classified
with the 2dFDR pipeline \citep{b+g99} and {\sc autoz}
redshifting code \citep{croom01}.

The main difference between reduced SDSS and 2dF spectra
is the lack of flux calibration for 2dF sources. An average flux
calibration for the 300B grating has been calculated by
\citet{lewis02} as part of the 2dF Galaxy Redshift Survey;
in our analysis we apply this correction to the spectra.

\subsection{Analysis}

The size of our sample is such that we do not manually inspect the
\civ\ line in each spectrum. Instead we develop an automated routine for
measuring the \civ\ line width in our sample. We do not include a
long discussion here (details are given in
appendix~\ref{sec:fitting}), but will state that we have developed
a line fitting routine that both measures the 50\,\%\ inter-percentile
velocity (IPV) width of the \civ\ line accurately and returns an
accurate error for that measurement. Our procedure is found to be
robust for spectral $\sn>3$\,\AA\pmo\ (observed frame), and so in the
analysis that follows this S/N cut is applied to our data.

\subsection{Final sample}

To appear in our final sample a spectrum must: be at the
right redshift to have the \civ\ line, have $\sn>3$\,\AA\pmo, and pass our
BAL tests.

To have the \civ\ line in the spectrum, and enough
surrounding coverage for the continuum fit, requires a redshift of $>$1.5
and $>$1.6 for 2dF and SDSS spectra respectively. At high redshift we impose a
further redshift limit of $z<3.3$. Beyond $z\sim3.3$ the \civ\ line
becomes mingled with the strong sky emission lines at the red end of
optical spectra. While the sky subtraction in our sample is generally
very good, residual correlated features can produce spurious results.

A significant proportion ($>10$\,\%) of \civ\ lines in QSO spectra
are effected by
strong broad-absorption features in their blue wing
(e.g. \citealt{tru06}). We develop an automated routine for finding
BALs in spectra, details of this routine are given in
appendix~\ref{sec:bal}, here we will simply state that overall it
removes $\sim35$\,\% of our spectra from the final analysis.

All of these limits result in a final sample of 13,776
line measurements that we consider reliable and use in the following
analysis. The line width results for this final sample can be found on
the 2SLAQ website ({\sc www.2slaq.info}).

\section{Results} \label{sec:results}

Before presenting the results of our line fitting we should comment on
the redshift distribution of our sample. 
The 2QZ, 2SLAQ and SDSS QSO samples are all selected based on their
optical colours. At lower redshifts the UV excess technique
(e.g. \citealt{s+g83}) or a variant is able to distinguish QSOs as `bluer' than
stars. This becomes ineffective for higher redshifts when the Lyman
break enters the $U$-band, reddening the colour of QSO. At redshifts
$\sim2.5-3$ QSO optical colours become intermingled in the stellar
locus and target selection is very incomplete over this range of redshifts
\citep{rich02b}. Beyond $z\sim2.5$, dropout techniques can be used to find high
redshift targets due to their lack of blue-UV flux.

\begin{figure}
\centering
\centerline{\psfig{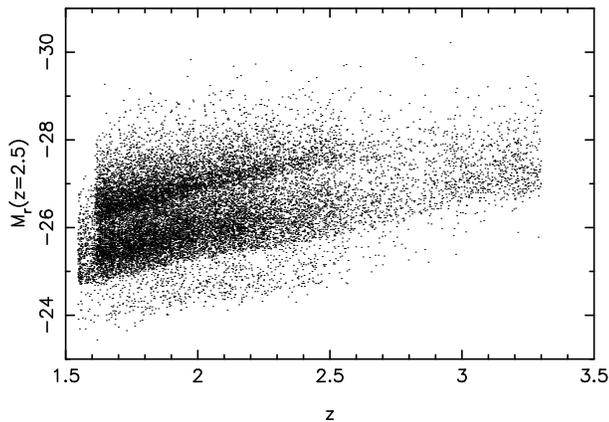}}
\caption{The absolute magnitude-redshift distribution for all objects
with an accepted fit to the \civ\ line.}
\label{fig_M_z_c}
\end{figure}

The result of these selection problems is a very uneven redshift
distribution for high-$z$ QSOs.
Fig.~\ref{fig_M_z_c} shows the magnitude-redshift
distribution of objects with a \civ\ line measurement. As
can be seen, the vast majority of objects are found in the main surveys
at redshifts less than $\sim2.5$ and there is a sparse number of
high-$z$ selected SDSS QSOs up to $z=3.3$. In addition, the irregular
magnitude distribution evident in Fig.~\ref{fig_M_z_c} is caused by
the differing magnitude limits of the surveys we draw our data from.

In Fig.~\ref{fig_M_z_c} we use $r$-band magnitudes since shorter
wavelength band passes will be affected by the Lyman break at the
redshifts we are sampling. We $K$-correct 
these using the SDSS QSO composite of \citet{van01}. The 2QZ catalogue
has data for the photographic $r$-band magnitude of objects as opposed to
SDSS $r$ \citep{fuk96}. However, the $r$ band photometry is incomplete
as 2QZ QSO candidates could be
selected without an $r$-band detection. For consistency we use the
$\bj$ magnitudes from the 2QZ catalogue. After
$K$-correcting the $\bj$ magnitudes we use a constant colour
correction calculated from the \citet{van01} template to transform to
the SDSS $r$-band. We note that there are very few 2QZ
QSOs beyond redshift three which is roughly the point at which
the Lyman break enters the $\bj$ band.

Following previous authors (e.g. \citealt{rich06b}), rather than
normalising the $K$-corrections to $z=0$ we use a redshift that
is more representative of our data and
removes systematic errors arising from the
large extrapolation to $z=0$. We choose $z=2.5$ as the zero-point of
our $K$-corrections, this amounts to a constant offset
of $M_r(z=0)-M_r(z=2.5)=0.36$. At $z=2.5$ we correct our $\bj$
magnitudes by $\bj-r=0.02$.

\subsection{Luminosity vs. line width}

\begin{figure}
\centering
\centerline{\psfig{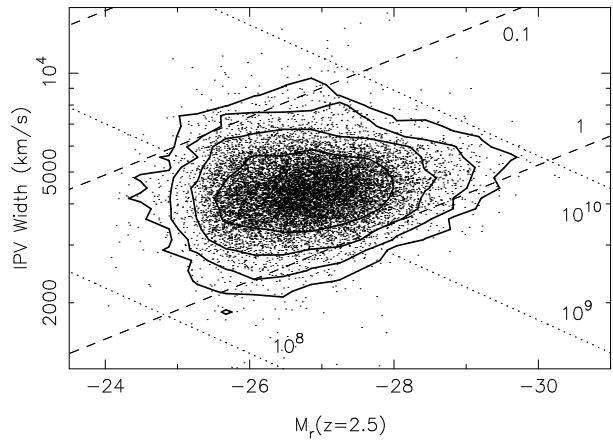}}
\caption{The results of our fitting of the \civ\
line. We plot the absolute $r$-band magnitude of the source vs. the
measured IPV width of the \civ\ line. Over-plotted are lines of
constant SMBH mass (dotted) and Eddington ratio (dashed) as a guide
to where these objects fall in mass-accretion space. Masses are
labelled in units of \msun.}
\label{fig_M_lw_c}
\end{figure}

Fig.~\ref{fig_M_lw_c} shows our sample's distribution in absolute
magnitude-line width space.
Contours on the plot are equally spaced in terms of the log of
the density of points. As a rough
guide to how these measurements would convert to SMBH mass and
accretion efficiency we have added to the diagram lines of constant
SMBH mass (dotted) and Eddington ratio (dashed). These were calculated
assuming the \citet{v+p06} virial mass calibration.

The distribution appears to be
constrained along lines of constant SMBH mass and Eddington ratio
such that the line at $\mbh=10^{10}$\,\msun\ defines the top of the
distribution. The lower limit appears to be at Eddington ratios
at or around one, indicating very little super-Eddington accretion.

Fig.~\ref{fig_M_lw_c} can be directly compared with Fig.~5 in
\citet{me2} which shows equivalent results for the \mgii\ line. The
two plots are qualitatively very similar. There is some indication of
an offset between the two distributions in SMBH mass-accretion
efficiency space with the \civ\ distribution tending towards slightly
larger Eddington ratios and lower SMBH masses.
However, one must keep in mind that the normalisation of these lines
is uncertain, potentially by as much as 0.5\,dex.
The differing zero-points in the
\mgii\ and \civ\ virial calibrations and/or systematics in the way
the lines in Fig.~\ref{fig_M_lw_c} and in \citet{me2} were derived
could explain the difference between the figures.



\subsection{Direct comparison with \mgii\ line widths from \citet{me2}}

\begin{figure}
\centering
\centerline{\psfig{file=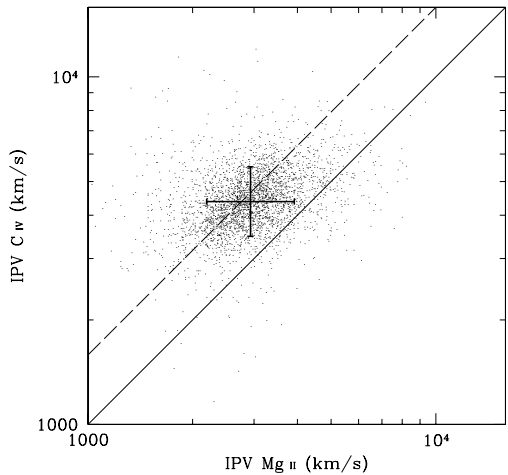,width=7.0cm}}
\caption{The IPV width of the \mgii\ line (taken from \citealt{me2}) plotted
against that of \civ\ for the 3197 objects with a measurement of both features.
The solid line shows the 1:1 relation while the dashed line
shows the expected ratio between the lines assuming
equation~\ref{equ_ipot_r}. The cross indicates the mean and rms
of the distribution.}
\label{fig_mg_civ}
\end{figure}

QSOs with redshifts between 1.5 and 2.3 will have both \civ\ and
\mgii\ in their optical spectrum, and can be used to make direct
comparisons between the two lines. The requirement to have a spectral
window around the emission line for continuum fitting means that the final
sample of QSOs with measurements for both the \civ\ and \mgii\
lines is limited to a redshift range $1.6<z<2.0$. In this range we
have 3197 spectra that have measurements of both the
emission lines; here and in \citet{me2}. Fig.~\ref{fig_mg_civ}
compares the velocity width
of the \mgii\ and \civ\ lines for these objects.

Two aspects of Fig.~\ref{fig_mg_civ} are of particular interest to
this work. 1) there does not appear to be an obvious
correlation between the widths of the two lines, and 2) there is
a clear offset such that the mean of the distribution does not lie on the
1:1 line. We discuss each of these points below.

\subsubsection{Reasons for offset}

The ionisation potential of \civ\ is $\sim6$ times that of \mgii. So
we might expect it to be emitted from a region closer to the
continuum source where the ionisation parameter is
higher. Stratification of emission regions is a
natural part of some BLR models (e.g. \citealt{bal95}), and is
borne out observationally by reverberation mapping
(e.g. \citealt{o+p02}).

If we take the oversimplified case in which the ionisation parameter,
$U$ ($U \propto F/n_H$; where $F$ is the ionising flux and $n_H$
is the gas density), required to ionise the \mgii\ and \civ\ BLRs
is proportional to their ionisation potential, $\chi$, then
\begin{equation}
\frac{U_{\rm \cive}}{U_{\rm \mgiie}}=
\frac{\chi_{\rm \cive}}{\chi_{\rm \mgiie}}.
\end{equation}
If we assume the flux follows an inverse square law and that the gas
density is approximately constant, and we assume that the velocity
field is dominated by virial motion then
\begin{equation}
F\propto r^{-2}, \hspace{0.5cm} n_H \sim const. \hspace{0.5cm} {\rm and} \hspace{0.5cm} r\propto v^{-2} \\
\end{equation}
which implies that
\begin{equation}
\left(\frac{v_{\text{\civ}}}{v_{\text{\mgii}}}\right)^4 =
\frac{\chi_{\text{\civ}}}{\chi_{\text{\mgii}}}.
\label{equ_ipot_r}
\end{equation}
%

While this model is clearly an oversimplification of the BLR it gives
the prediction that the ratio between the line widths of \mgii\ and
\civ\ should be the fourth root of their ratio in ionisation potential,
or a factor of $\sim1.58$. The dashed line in
Fig.~\ref{fig_mg_civ} shows the $v_{\rm \cive}=1.58v_{\rm \mgiie}$
relation. The dashed line agrees remarkably well with the measured
zero-point of the distribution, and indicates why we may expect an
offset in the observed distribution. On the other hand, the
scatter around the zero-point demonstrates that there is no simple way to
relate the width of the \mgii\ and \civ\ lines.

\subsubsection{The \civ-\mgii\ correlation}

The widths of \mgii\ and \hb\ have been shown to correlate
well for QSOs \citep{m+j02,sal07} and we might expect a similar
correlation for \civ. However, Fig.~\ref{fig_mg_civ} does not display
a clear correlation between \mgii\ and \civ\ line width (see also
\citealt{shen08}).

If we perform a Spearman rank test on our data we find a
significant correlation ($r_s=0.35$; $P(r_s)\ll0.01$). On the
other hand, we find 0.09\,dex scatter around a y-on-x best
fit to the distribution, hardly reducing the 0.1\,dex scatter in the
original data.

If the average ratio between \mgii\ and \civ\ line width changes with
luminosity or redshift, any correlation in Fig.~\ref{fig_mg_civ} would
be blurred by our inclusion of QSOs with a range of these
properties. As a test we bin our sample by luminosity and
redshift, and recalculate the Spearman rank coefficient in each
bin. The sample of objects that have both \civ\ and \mgii\ in their
spectra span a relatively small redshift range, and so we divide the
sample into two redshift bins by the approximate midpoint at
$z=1.8$. We then also divide the sample into half-magnitude
bins. Table~\ref{tab:cor_coef_m} shows the number of objects in each
bin and the calculated $r_s$. It is clear that $r_s$ increases
somewhat with luminosity, and stays roughly constant with
redshift. The zero-point to the relation stays almost constant with
respect to luminosity, varying by less than 0.01\,dex between the the
faintest and brightest bins.

\begin{table}
\begin{center}
\caption{These data look at the correlation between \civ\ and \mgii\
  line widths when our sample is binned by luminosity and redshift. We
  give the magnitude limits of each luminosity bin as well as the
  number of objects in each bin and the Spearman rank
  coefficient. These results are shown for
  objects in two redshift bins as well as the whole sample.}
\label{tab:cor_coef_m}
\begin{tabular}{ccccccc}
\hline \hline
  & \multicolumn{2}{c}{$z<1.8$} & \multicolumn{2}{c}{$z>1.8$} &
  \multicolumn{2}{c}{All} \\
$M_r(z=2.5)$ range & $N$ & $r_s$ & $N$ & $r_s$ & $N$ & $r_s$ \\
\hline
$M>-26$       & 475 & .24 & 152 & .27 & 627 & .24 \\
$-26>M>-26.5$ & 471 & .32 & 164 & .35 & 636 & .33 \\
$-26.5>M>-27$ & 614 & .34 & 351 & .32 & 966 & .33 \\
$-27>M>-27.5$ & 275 & .47 & 270 & .39 & 546 & .43 \\
$-27.5>M>-28$ & 122 & .49 & 143 & .30 & 265 & .39 \\
$-28>M$       & 46  & .39 & 73  & .44 & 119 & .41 \\
\hline \hline
\end{tabular}
\end{center}
\end{table}

It is likely that reduced measurement error is the cause of the
improving correlation (increasing $r_s$) with increasing
luminosity. In table~\ref{tab:cor_coef_sn} we bin our sample by S/N
rather that luminosity or redshift. Table~\ref{tab:cor_coef_sn} shows
that the correlation improves for objects with higher S/N spectra. We
also give the mean percentage error on the \civ\ line width measurements in
each bin (\civ\ line widths invariably have larger measurement errors;
see section~\ref{sec_civ_mg_disp_comp}), and the rms scatter around the
y-on-x least-squares fit.


\begin{table}
\begin{center}
\caption{These data look at the correlation between \civ\ and \mgii\
  line widths when our sample is binned by S/N. The first column gives
  the S/N range of the bin. Also given is the number of objects, the
  Spearman rank coefficient, the rms scatter around a y-on-x least
  squares fit, and the mean percentage error on the line width
  measurements in each bin.}
\label{tab:cor_coef_sn}
\begin{tabular}{ccccc}
\hline \hline
 & & & rms around & Mean \\
S/N\,\AA\pmo\ range & $N$ & $r_s$ & best fit (dex) & error (\%) \\
\hline
$3<\rm S/N<5$   & 238 & .22 & 0.129 & 21 \\
$5<\rm S/N<8$   & 564 & .25 & 0.104 & 16 \\
$8<\rm S/N<13$  & 909 & .35 & 0.087 & 10 \\
$13<\rm S/N<20$ & 876 & .42 & 0.073 & 7  \\
$20<\rm S/N$    & 522 & .42 & 0.082 & 5  \\
\hline \hline
\end{tabular}
\end{center}
\end{table}

\begin{figure}
\centering
\centerline{\psfig{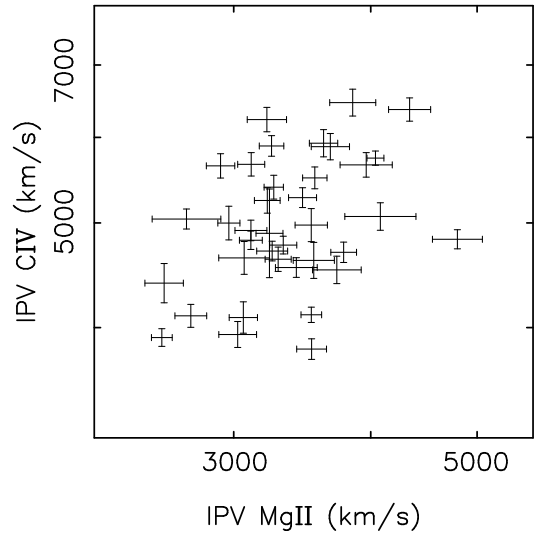}}
\caption{The correlation between \mgii\ and \civ\ line widths in the
  highest S/N ($>$40\,\AA\pmo) spectra in our sample.}
\label{fig_mg_civ_hsn}
\end{figure}

Table~\ref{tab:cor_coef_sn} shows that the correlation between \mgii\
and \civ\ line widths may be somewhat better than shown in
Fig.~\ref{fig_mg_civ}. Fig.~\ref{fig_mg_civ_hsn} shows the relation
between \mgii\ and \civ\ line widths for the 37 spectra in our sample
that have both lines and $\sn>40$\,\AA\pmo. The correlation in
Fig.~\ref{fig_mg_civ_hsn} is, perhaps, more apparent than
Fig.~\ref{fig_mg_civ}. However, a Spearman rank test implies the
correlation is only marginally significant ($r_s=0.36$;
$P(r_s)=0.03$). The rms scatter around a y-on-x regression line is
0.055\,dex compared to 0.060\,dex raw scatter in \civ\ line widths.
We also measure the intrinsic scatter in the data,
accounting for measurement error, around a best-fit that minimises the
2D $\chi^2$ and find it to be 0.055\,dex.

To summarise, these results show that there is a significant
correlation between the widths of \mgii\ and \civ\ lines in our
sample. However, the correlation is weak, in that the dynamic range of
line widths is only marginally broader than the intrinsic scatter
around a best-fit regression line.

There are many potential sources of intrinsic scatter in
Fig.~\ref{fig_mg_civ}. As we have seen, the higher ionisation
potential of \civ\ may indicate that its emission arises in a
smaller part of the BLR, closer to the SMBH. It has been suggested that the
emission regions for high ionisation lines may differ dynamically
from that of lower ionisation lines \citep{rich02,elvis04}.
There is, therefore, some concern as to how well the \civ\ virial mass
estimators would agree with those for \hb\ and \mgii\ (see discussion
in section~\ref{sec:vir_wrap}).

\section{Dispersion in the line width distribution} \label{sec:disp}

Motivated by results presented in \citet{me2} we
examine the second moment of the \civ\ line width
distribution; in particular, we are interested in how it changes with
QSO luminosity. In brief, the analysis of \citet{me2}
is performed in three steps:

\begin{itemize}
\item{Bin our sample by redshift and luminosity}
\item{Calculate the 68.3\,\% interpercentile range of the line
widths to characterise the dispersion in each bin.}
\item{Correct the dispersion with the median error on the line
widths in each bin by equation~7 in \citet{me2} to
take account of scatter in the data due to measurement error.}
\end{itemize}

We find that the dispersion results for \civ\ are
considerably more sensitive to the methods used in their derivation
than was the case in \cite{me2} for \mgii. The
increased sensitivity to the fitting procedure is due to two
compounding issues. Firstly, since the
process of measuring the width of the \civ\ line is more complicated
than for \mgii, the resulting errors on these widths tend to be
larger. Secondly, we find less
intrinsic dispersion in \civ\ line widths than was found for
\mgii. A narrower intrinsic line width distribution, combined with
larger measurement errors, makes deconvolving their separate
effects on the measured line width distribution more difficult.

\subsection{What affects the dispersion in the measured IPV width
distribution?}

The measured IPV width of a line and its associated error depend
on the spectral window in which the IPV width is calculated. We find
that the way this window is defined can affect the results we derive
for the dispersion in the IPV width distribution.


When analysing the \mgii\ line, \citet{me2} defined a
region in the spectrum between $\pm1.5$ times the FWHM of a Gaussian
fit to the line within which they calculated the IPV width. We find
that adopting a fixed region between 1475 and 1625\,\AA, within which
to calculate the IPV width, gives more stable results for the \civ\
line. We choose this region as being wide enough to enclose $>$99\,\%
of the flux for average \civ\ lines and $>$90\,\% for the broadest 
lines. We do not make the limits wider as they would then
encroach on the region used for the continuum fit. Here we discuss
the effect of using a fixed region for our IPV calculations

%

\begin{figure*}
\centering
\centerline{\psfig{file=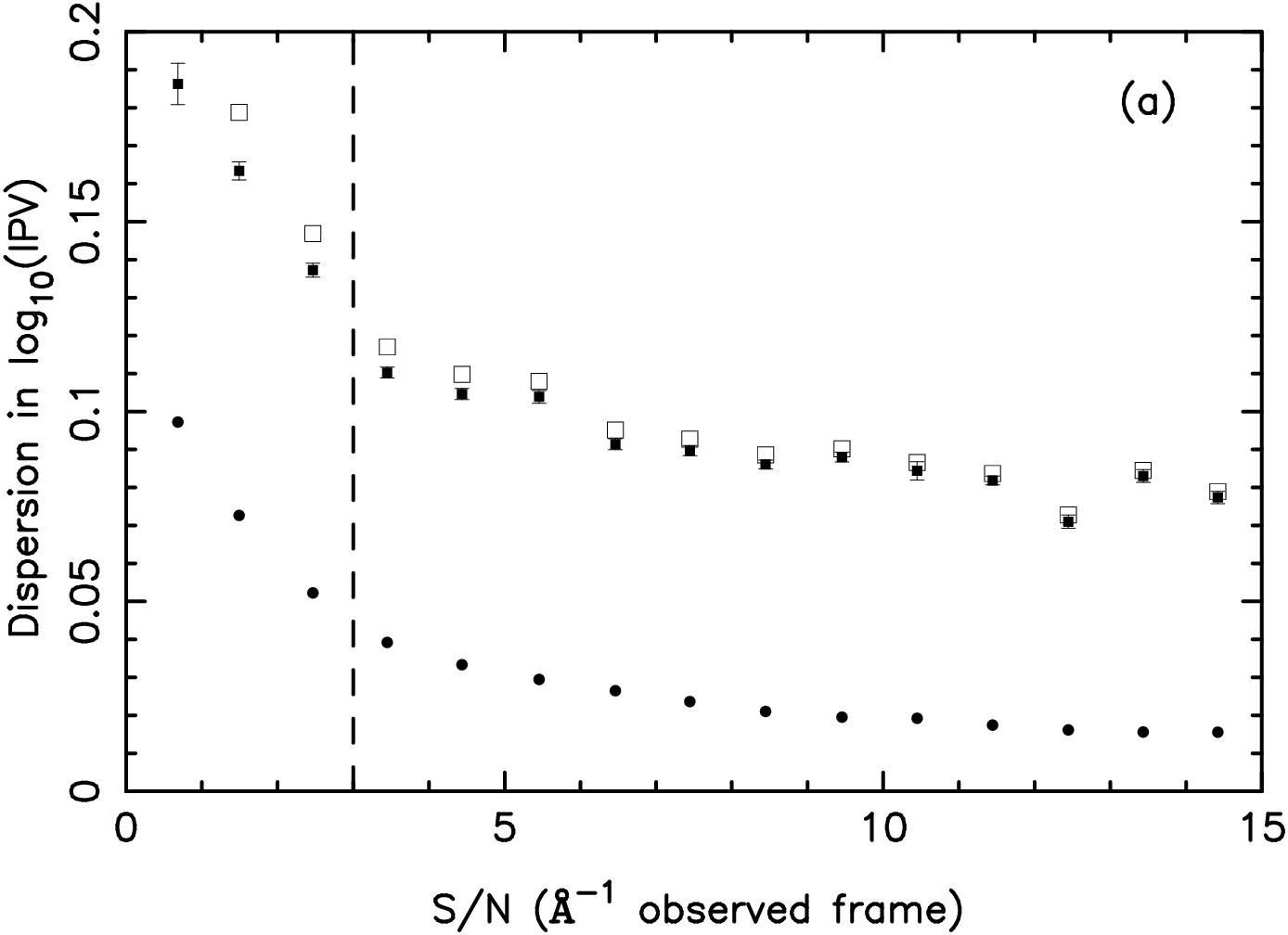,width=7.5cm,angle=-90}\hspace{0.5cm}\psfig{file=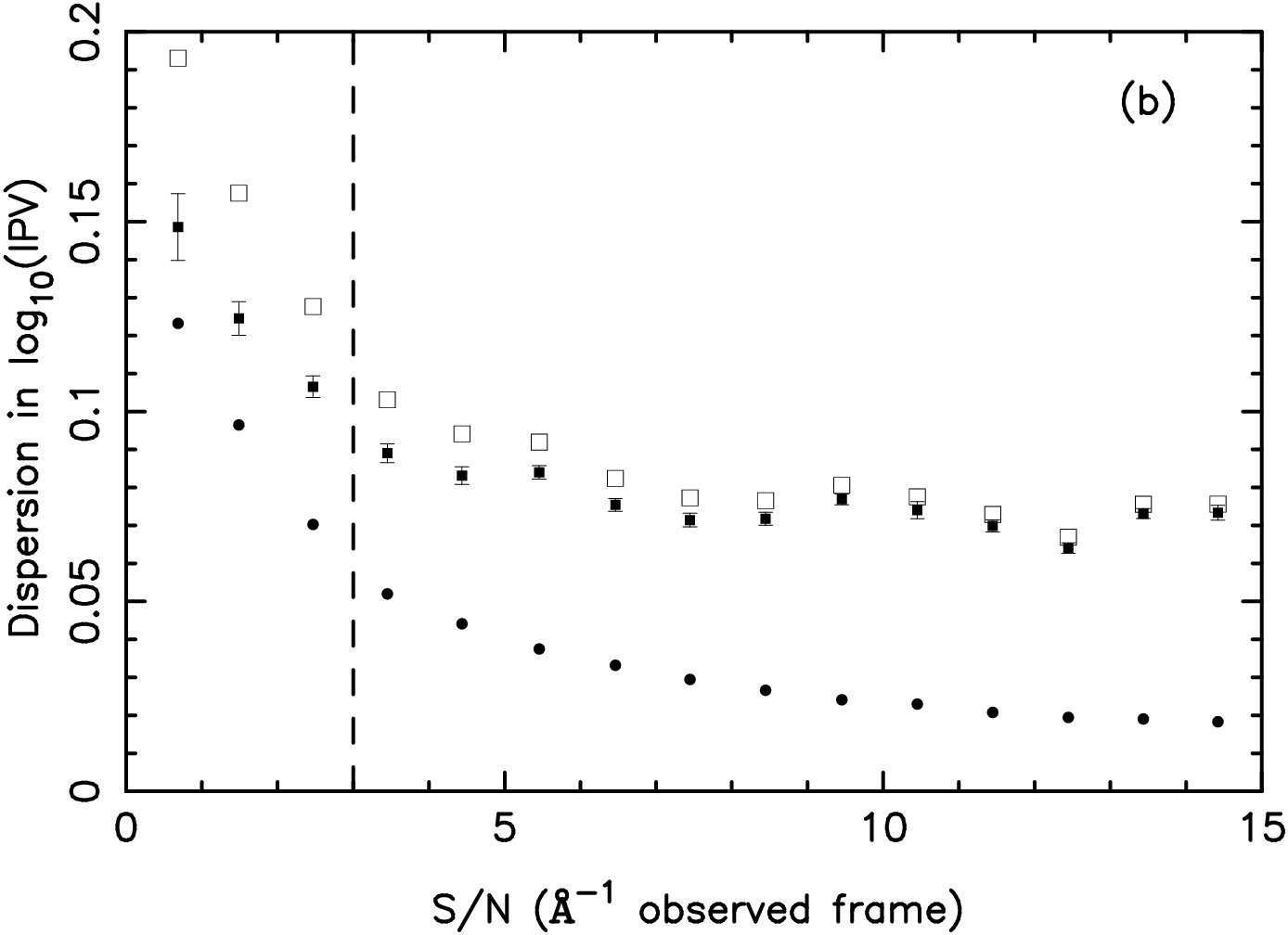,width=7.5cm,angle=-90}}
\caption{The dispersion in \civ\ IPV width as a function of spectral
S/N. In \empha\ the IPV width is calculated over a spectral window
defined by the Gaussian fit to the line. In \emphb\ the window has a
fixed range. The raw dispersion in IPV widths is plotted as open
squares, this is corrected by the median error (filled circles) to
estimate the intrinsic dispersion in \civ\ line widths (filled
squares). The dashed line shows S/N$=$3 below which we do not have
confidence in our fitting. Above S/N$=$3 the fits in \empha\ show
slightly more dependence on S/N than in \emphb.}
\label{fig_civ_sn_disp}
\end{figure*}

In Fig.~\ref{fig_civ_sn_disp} we show the dispersion in
IPV width as a function of S/N. We do not distinguish between 
2dF and SDSS spectra since we find no difference between them when
compared at the same S/N. Fig.~\ref{fig_civ_sn_disp}\empha\
shows the behaviour when the IPV width is measured
over a spectral region defined by the Gaussian fit to the \civ\ line, while
the widths in \emphb\ are calculated in a fixed spectral window. In
each plot the measured dispersion in IPV widths is shown as the open
squares; this is corrected by the median error (shown as circles) to
give an estimate of the intrinsic dispersion in \civ\ line widths
(shown as filled squares with error bars). The dashed line shows
S/N=3\,\AA\pmo, below which we know our line width measurements are
biased from simulating low S/N spectra.



We find that employing a fixed region for the IPV
width calculation
results in the derived intrinsic dispersion in IPV widths being less
dependent on the spectral S/N. The dispersion in IPV widths calculated
with a variable region tends to be greater at lower S/N and their errors
tend to be smaller than those calculated using a fixed spectral
window.

In general, the fixed window will be larger than a window defined
by the Gaussian fit, and one expects the errors on the IPV widths
calculated over a fixed window to be larger.
It is less clear why we find less scatter in IPV widths calculated
with a fixed spectral window when compared with those calculated over a
variable region. The indication is that using our Gaussian fits is adding
uncertainty to our IPV calculations in some manner that is not
reflected in their errors. The level of uncertainty
is very low ($<$0.01\,dex for S/N$\sim$4). It is possible that the
non-linear multi-Gaussian fits to the spectra are unreliable at this level.

To have confidence in our analysis
we need to identify what causes the increase in the corrected
dispersion in IPV widths calculated over a variable spectral window as
we go to lower S/N. Is this due to increased noise? Alternatively,
because of the
correlation between the intrinsic luminosity of a source and the
spectral S/N, is this due to an inherent property of the quasars in
our sample? i.e. is this increase due to a correlation between the intrinsic
dispersion in \civ\ line widths and quasar luminosity as was evident
for \mgii\ (see \citealt{me2})?

\begin{figure}
\centering
\centerline{\psfig{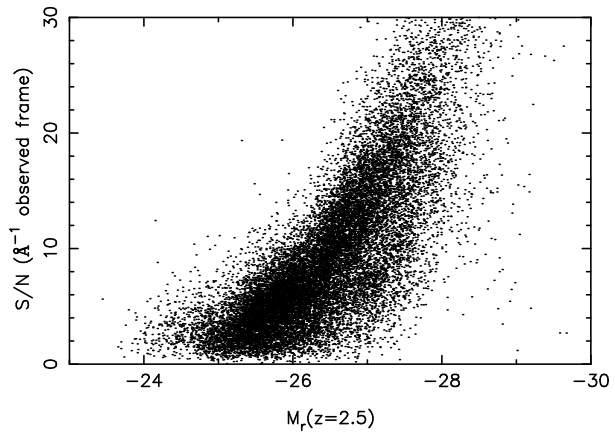}}
\caption{The relation between the spectral S/N calculated in the local
\civ\ region of the spectrum and the absolute magnitude of quasars in
our sample.}
\label{fig_civ_M_sn_disp}
\end{figure}

The S/N and absolute magnitude of the QSOs in our sample do correlate
(Fig.~\ref{fig_civ_M_sn_disp}). To test what is causing
the increase in the corrected dispersion in IPV widths calculated over
a variable spectral window as 
we move to lower S/N we add noise to high S/N spectra, calculate the IPV
width in this noisier spectrum and then replot the dispersion in IPV
widths as a function of S/N.

\begin{figure*}
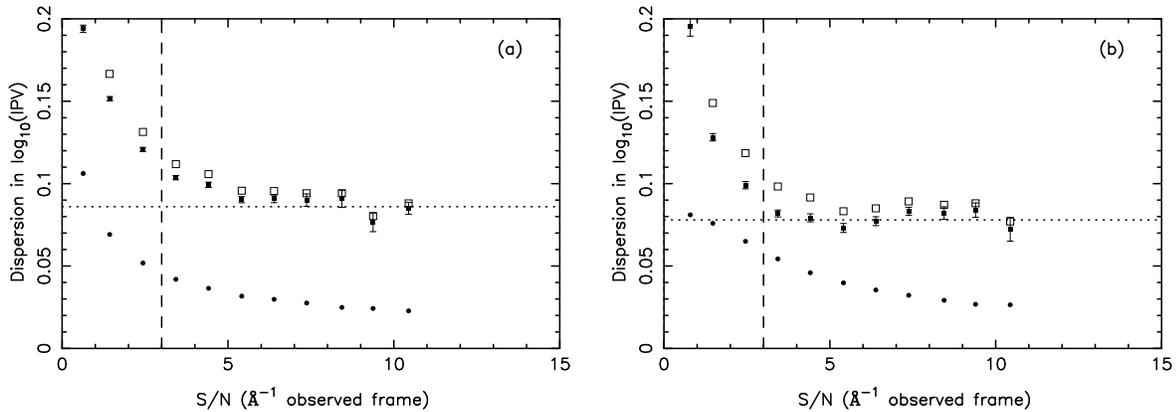

\centering
\centerline{\psfig{file=disp_sn_2degrade.ps,width=7.5cm,angle=-90}\hspace{0.5cm}\psfig{file=disp_sn_8degrade.ps,width=7.5cm,angle=-90}}
\caption{The dispersion in IPV line widths measured from spectra which
have had their S/N degraded by a factor of three. In \empha\ the IPV
width is calculated over a spectral window 
defined by the Gaussian fit to the line. In \emphb\ the window has a
fixed range. Symbols are as in Fig.~\ref{fig_civ_sn_disp}, the dashed
line shows the S/N$=$3 line below which we do not trust our fitting
results. The dotted line shows the dispersion in IPV widths for
objects with original spectra with S/N$>$9, and should line up with
degraded spectra with S/N$>$3.}
\label{fig_civ_sn_disp2}
\end{figure*}

We degrade every spectrum in our data
such that their resulting S/N is 1/3 its original value. In doing so we also
modify the error on each pixel to take account of this addition of
noise. Fig.~\ref{fig_civ_sn_disp2} shows the new dispersion
results as a function of S/N. When we add noise to the spectra
artificially we find an
almost identical relation between dispersion in IPV width and S/N as
was evident in Fig.~\ref{fig_civ_sn_disp}.

The similarity between Figs.~\ref{fig_civ_sn_disp} and~\ref{fig_civ_sn_disp2}
indicates that the increase in corrected dispersion
observed towards lower S/N is due to noise and not an intrinsic
property of the QSO emission lines. To further illustrate the effect
we average the
dispersion calculated with our original data for S/N$>$9 and plot this in
Fig.~\ref{fig_civ_sn_disp2} as the dotted line. For S/N$>$9 in
Fig.~\ref{fig_civ_sn_disp} the dispersion in IPV width is relatively
constant with S/N. A S/N of 9 in Fig.~\ref{fig_civ_sn_disp}
corresponds to S/N$=$3 in Fig.~\ref{fig_civ_sn_disp2}. Clearly the
points at S/N$>$3 follow the dotted line in
Fig.~\ref{fig_civ_sn_disp2}\emphb\ more closely than in \empha.

\begin{figure*}
\centering
\centerline{\psfig{file=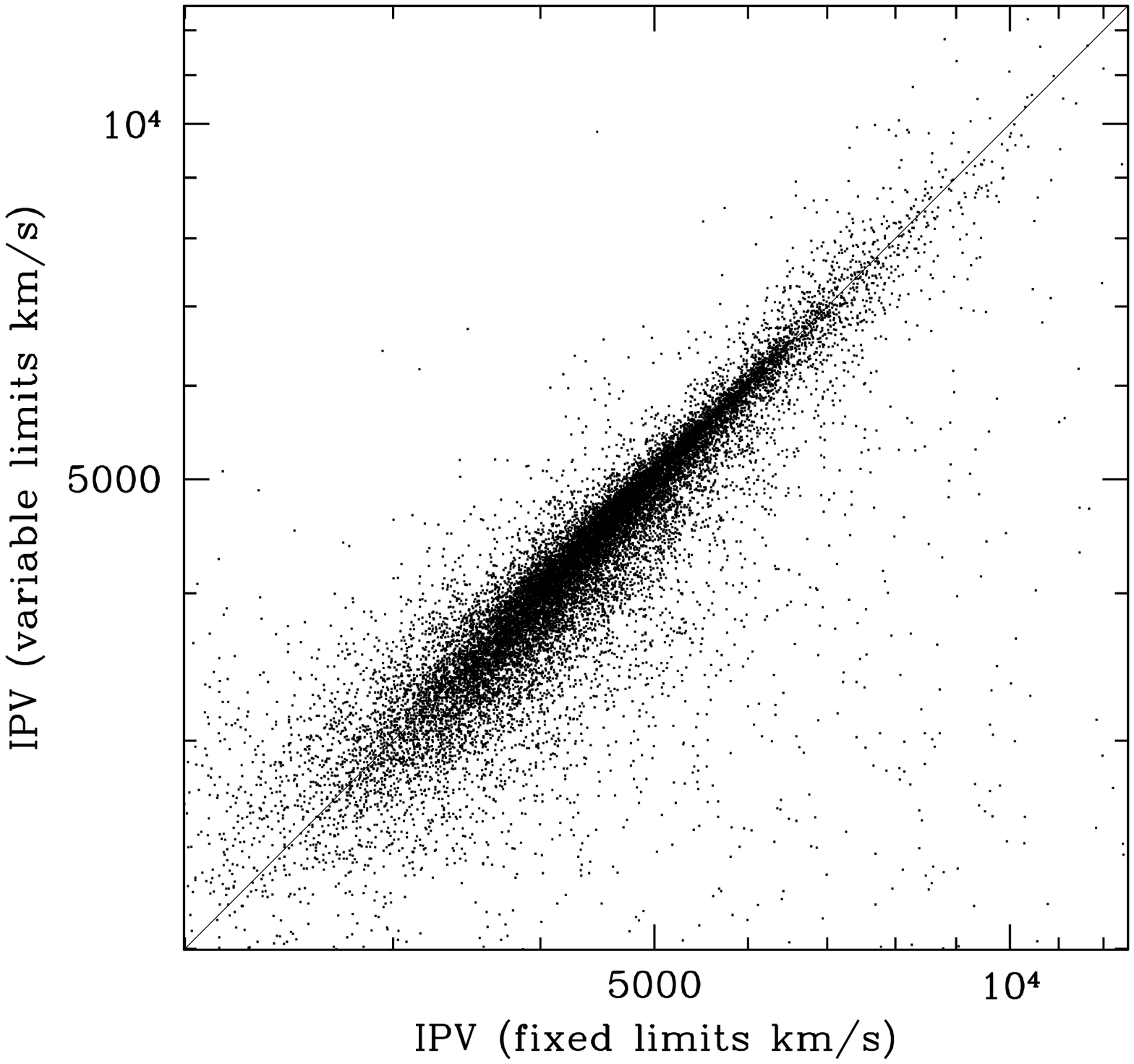,width=7.0cm}\psfig{file=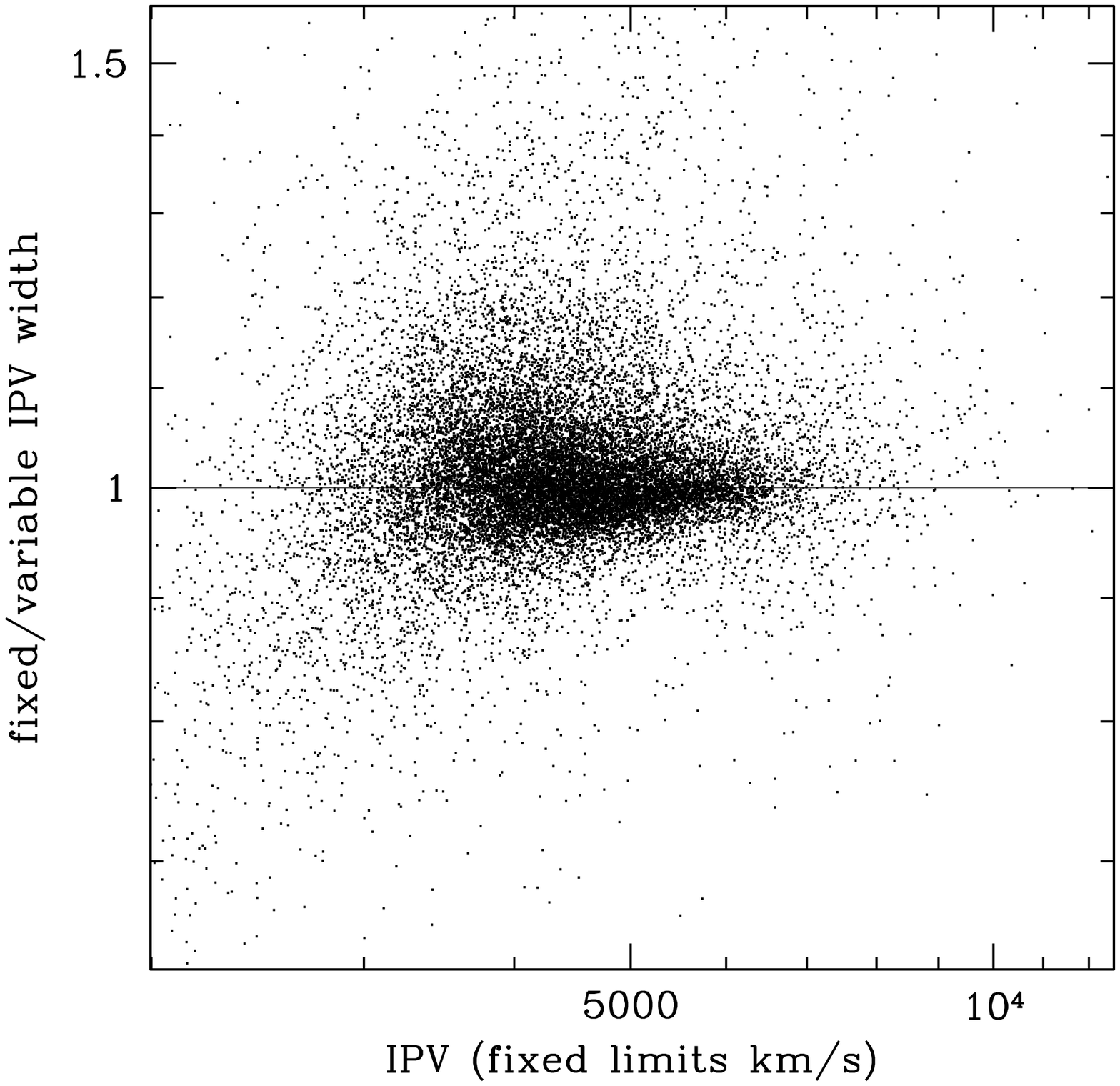,width=7.0cm}}
\caption{Comparisons between IPV widths calculated over a fixed
spectral window, and with a window which is defined by the Gaussian
fit. \empha\ shows the direct comparison and \emphb\ shows their ratio
as a function of line width.}
\label{fig_civ_fix_var_comp}
\end{figure*}

Fig.~\ref{fig_civ_fix_var_comp} compares the IPV line widths measured
over a fixed spectral window, and with a window which is defined by
the Gaussian fit. In Fig.~\ref{fig_civ_fix_var_comp}\empha\ we plot a
straight comparison between the measured widths and in \emphb\ the
ratio is plotted. There is scatter between the values, furthermore
this increases for narrower line widths where the difference between
the IPV windows will be greatest. For narrower lines the IPV widths
measured over a fixed range tend to be larger, although there is
scatter in both directions. Overall it does not seem that we are
biasing our results significantly by fixing the region over which we
calculate the IPV width.

Since the relation between the corrected dispersion in IPV widths and
S/N is flatter when using a fixed spectral window to calculate the IPV
widths, we prefer this method for our final fitting procedure. There remains
a slight trend in the corrected dispersion with S/N such that
the corrected dispersion increases by $\sim0.015$\,dex towards
S/N$=$3. We find that the trend can be altered depending on the statistics we
employ to define the dispersion in IPV widths and the average error
used in the correction.

In all of the above figures we have followed the method used in
\citet{me2}. We have taken the 68.3\,\% inter-quartile range
to parametrise the dispersion in the IPV width distribution, and used the
median error to correct to the intrinsic dispersion.

\begin{figure}
\centering
\centerline{\psfig{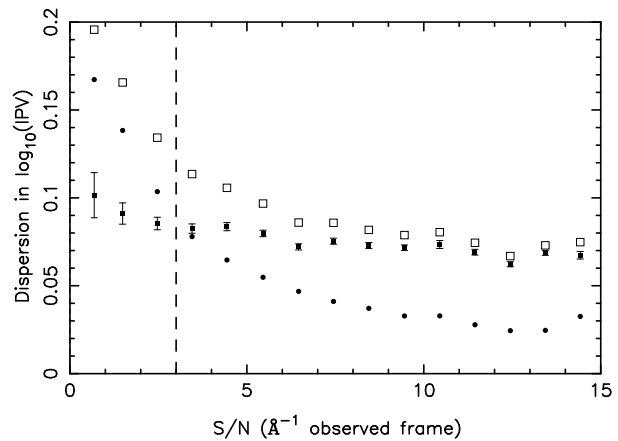}}
\caption{Here we show the measured dispersion in \civ\ line widths as
a function of S/N. Open squares show the raw dispersion measured as
the rms of the distribution. These are corrected by the mean error on
the line widths (circles) to estimate the intrinsic dispersion in line
widths (filled squares).}
\label{fig_civ_sn_disp3}
\end{figure}

Taking the rms of the IPV width distribution and correcting by
the mean error we find the results shown in
Fig.~\ref{fig_civ_sn_disp3}. Here the trend is less than in
Fig.~\ref{fig_civ_sn_disp}\emphb, although the effect is small for ${\rm
S/N}>3$. 
Nonetheless, in the analysis that follows, we employ the rms to
calculate dispersion and correct by the mean error of the IPV widths.

\subsection{Dispersion in IPV widths vs. luminosity and redshift}

\begin{figure}
\centering
\centerline{\psfig{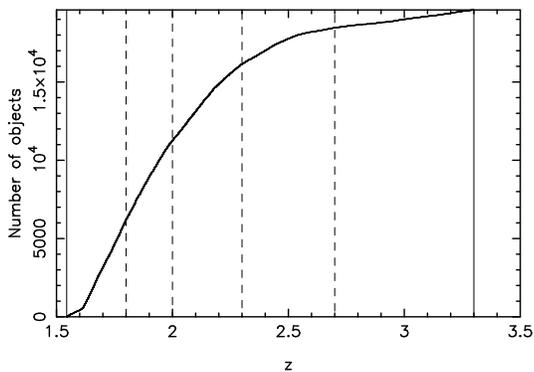}}
\caption{The cumulative redshift distribution of objects in our
sample, i.e. the y axis plots the number of objects with redshift less
than $z$. Vertical lines on the plot show the limits of the redshift
bins we will be using in our analysis.}
\label{fig_N_z_c}
\end{figure}

To characterise the dependence of the dispersion in \civ\ line width on
quasar luminosity and redshift we bin our data by $L$ and $z$ and
calculate the dispersion in each bin.
Due to the uneven redshift distribution of our sample
(Fig.~\ref{fig_M_z_c}) we do not choose evenly spaced redshift bins
for our analysis. Fig.~\ref{fig_N_z_c} shows the cumulative redshift
distribution of our sample over which we have marked the
limits of the redshift bins we will use. The three lowest redshift
bins have roughly equal width and equal numbers of objects.
The redshift bin centred at $z=2.5$ contains QSOs approximately during
the quasar epoch. Finally we have a high redshift bin 
of objects with $z>3$, potentially before the quasar epoch
\citep{rich06}.

\begin{figure}
\centering
\centerline{\psfig{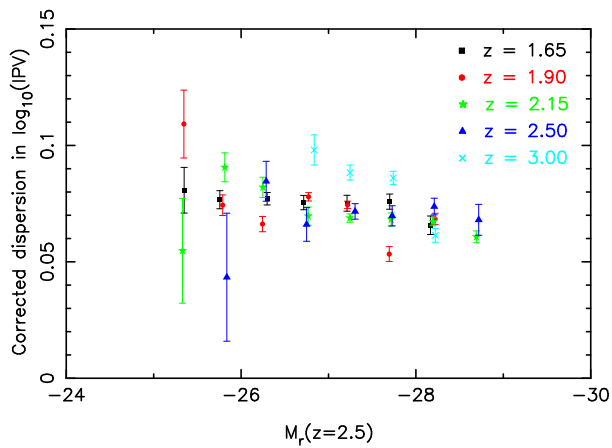}}
\caption{The dispersion in \civ\ line width in magnitude-redshift bins
as a function of magnitude. Each redshift bin is plotted with a
different colour and the midpoint of these bins is indicated on
the top right.}
\label{fig_M_disp_c}
\end{figure}

In each redshift bin we calculate the dispersion in \civ\ IPV line
width as a function of luminosity and plot the results in
Fig.~\ref{fig_M_disp_c}. We find no clear trend between QSO luminosity
and the dispersion in \civ\ IPV width. The lowest four redshift bins
appear to be equivalent, but there is an indication that the
highest redshift bin is offset to a higher dispersion. Furthermore,
there is a suggestion that the highest redshift bin shows an inverse
correlation between the dispersion in \civ\ line width and luminosity
although the dynamic range is small.

The evidence for increased dispersion at high redshift is
inconclusive; if the result is real it may be indicating that there
is more scatter in SMBH mass/Eddington ratio.
At such high redshifts the SMBH mass function can only be
steeper and we would expect to find many fewer SMBHs with
$\mbh>10^{10}$. It is hard to imagine a population of
objects capable of broadening the active SMBH mass distribution
towards higher masses at such an epoch. Alternatively, there could be
more super-Eddington accretion at high redshift. But again, it is
difficult to imagine a reason why the Eddington limit would be a weak
constraint at $z>2.6$ and then become a stronger limit at lower redshift.
As an alternative (non-virial) explanation, the \civ\ line width
could, potentially, be related to outflows from QSOs. The larger
dispersion in \civ\ line widths at high redshift may be indicating
that outflows were more common/more varied in the high redshift
Universe.

\subsection{Comparing the dispersion in \civ\ and \mgii\ line widths}
\label{sec_civ_mg_disp_comp} 

\begin{figure}
\centering
\centerline{\psfig{file=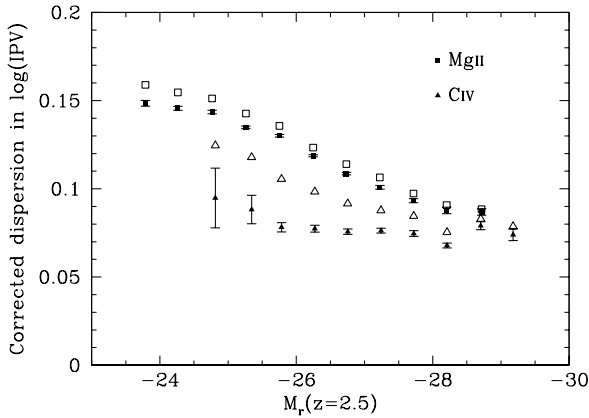,width=8.0cm,angle=-90}}
\caption{Comparison between the dispersion in \mgii\ (squares) and
\civ\ (triangles) line
widths. Each are plotted against the absolute $r$-band magnitude
$K$-corrected to $z=2.5$. Open symbols show the raw dispersion in the
data calculated as the rms of the line width distribution. Solid
symbols give the implied intrinsic dispersions once we have corrected
by the mean of the errors on the IPV width measurements.}
\label{fig_M_disp_cm}
\end{figure}

\citet{me2} showed that the dispersion in \mgii\
line width depends on QSO luminosity, but Fig.~\ref{fig_M_disp_c}
shows no clear signs for such a dependence for \civ. In
Fig.~\ref{fig_M_disp_cm} we take the \mgii\ line widths from
\citet{me2} and make a direct comparison between the
dispersion results for \civ\ and \mgii.

In Fig.~\ref{fig_M_disp_cm} we calculate the intrinsic dispersion in
the line width distributions in the same way for both \civ\ and
\mgii. That is we calculate the 
rms of the IPV width distribution (open symbols) and correct this by
the mean of the errors on the IPV width in each bin to give the
intrinsic dispersion in line width (shown as solid symbols).
There is a clear offset between the \mgii\ and \civ\ data. The
uncorrected dispersion in IPV width is larger for \mgii\ at all but
the brightest luminosities. In addition, the errors on the \mgii\ line
widths are smaller.


The width of the \mgii\ line can be measured with more precision than
\civ\ for two reasons.
Firstly, in the case of \civ\ we fit a linear continuum between two
45\,\AA\ wide windows either side of the \civ\ line. We are forced
into using small windows due to emission from other (sometimes
unknown) ions in the \civ\ line region. In the case of \mgii\
a much wider ($>450$\,\AA) spectral window can be used in the
continuum fit. This results in a considerably
more precise model for the local emission.


In addition to a more precise continuum fit, \mgii\ does not have
local contaminating emission (apart from iron emission that is well
fit by a single template)
which makes the \civ\ fitting more difficult. The \heii\ and \oiii\
emission lines on the red wing of \civ\ add further uncertainty into
the line width calculations.

The differing behaviour between \civ\ and \mgii\ in
Fig.~\ref{fig_M_disp_cm}, combined with the lack of a strong
correlation between these lines (Fig.~\ref{fig_mg_civ}), lends further
weight to the argument that the emission regions for these two lines
are distinct.

\section{Virial SMBH mass estimation with \mgii\ and \civ} \label{sec:vir_wrap}

Above we have compared \civ\ and \mgii\ line widths measured in spectra
which have both lines, and find considerable scatter in the
comparison. We find that the dispersion in \civ\ line width is smaller
than that for \mgii\ and does not show the same trend with luminosity
that \mgii\ exhibits.

Overall, the differing behaviour of the \mgii\ and \civ\ lines is a
concern when employing these as virial SMBH mass indicators. The lack
of a clear correlation between the widths of the two lines may be of most
concern since, for a virialised BLR, these should correlate
well. However, even without a correlation between \mgii\ and \civ\ line
widths, virial SMBH estimations using the two lines are still
consistent.

The consistency occurs because the dynamic range in line widths
for both \civ\ and \mgii\ is less than $\sim0.15$\,dex, or
$\sim0.3$\,dex in SMBH mass (under the virial assumption). The quoted
uncertainty on virial SMBH mass estimates is typically
larger than $\sim0.3$\,dex (e.g. $0.32$\,dex for the \civ\ 
calibration from \citealt{v+p06}, and $0.33$\,dex for \mgii\ from
\citealt{m+d04}). Hence the line width term in virial mass estimators
has only a weak effect on estimates for SMBH masses: The luminosity
term is responsible for the dynamic range of virial SMBH mass
estimates. Therefore, virial SMBH mass estimates from different lines
correlate, even if the widths of the emission lines themselves show no
correlation. If virial mass calibrations can not estimate SMBH masses
to a higher degree of accuracy than the dynamic range in line width,
the usefulness of the line width is unclear. Instead, virial
estimators may only appear to work due to their luminosity dependence.

While many of our results imply there are problems with the virial
technique for estimating SMBH masses, we also find evidence supporting
the virial assumption. We find that the offset between the average
\civ\ and \mgii\ line width is consistent with a simplistic model which
assumes a photoionised BLR with virial velocities; potentially this
indicates that both the high and low ionisation BLR are virialised and
so could be used to estimate SMBH masses. However, scatter around this
mean relation shows that this simple interpretation alone is inadequate.

It is unclear how all of our results can be accommodated in a single
model for the BLR, and what the eventual impact will be on the virial
technique. However, it is clear that care needs to be taken when using
virial SMBH mass estimates, and a better understanding of the BLR is
necessary before virial mass estimates can be considered to be
unbiased.

\section{The geometry of the \civ\ BLR} \label{sec:geometry}

If the \civ\
BLR velocity field is in any way asymmetric then the width of the \civ\
line will depend on the viewing angle. We find
only a very small level of dispersion in \civ\ line widths at all
luminosities, and use this to constrain models for the velocity field of
the BLR.

Perhaps the most common toy model for the BLR consists of a component
confined to a disk (either rotating or as a wind) as well as a random
isotropic component. Various parameterisations for this model can be
found in the literature (e.g. \citealt{j+m06,col06,lab06,me2}). In the
following discussion we will use
\begin{equation}
\label{equ:disk_mod1}
{\rm line\,width}\approx \sqrt{v_d^2\sin^2(\theta)/2 + v_r^2/3}
\end{equation}
where $v_d$ and $v_r$ are the disk and random velocities
respectively. This model assumes that the disk and random components
have approximately Gaussian velocity profiles, and that the emitting
regions are not distinct. The factors of 2 and 3 occur due to the
dimensional  confinement of the two components.

We are not interested in absolute line
width measurements, only the effect of orientation. We can therefore
transform equation~\ref{equ:disk_mod1} into a function of a single variable
that describes how disk-like (or not) the BLR is. To this end we must
tie $v_d$ and $v_r$ together under some assumption. We assume that,
viewed edge on (i.e. at $\theta=90^\circ$), the disk and random components are
indistinguishable. That is $v_d^2/2$$+$$v_r^2/3$$=$$const$.
We then define $a$ by
\begin{equation}
a^2=\frac{v_d^2/2}{v_d^2/2+v_r^2/3}.
\end{equation}
Hence if $a=1$ the BLR is disk-like and if $a=0$ it is spherically
symmetric. We then have
\begin{equation}
\label{equ_par_d2}
{\rm line\,width}\propto \sqrt{a^2\sin^2(\theta) + (1-a^2)}
\end{equation}
as our parametrisation for the BLR.

Given our simple model we have defined the effect of orientation on
line width and,
assuming an opening angle to QSOs defined by the extent of their
molecular torus (see Fig.~13 in \citealt{me2}), we can calculate the
dispersion in line width due to orientation effects.
Fig.~\ref{fig_M_disp_cm} shows that there is $\sim$0.08\,dex scatter in
\civ\ line width. We use this value to calculate an upper limit for $a$
depending on the assumed opening angle to QSOs. That is the maximum
value $a$ can take to be consistent with our data.

\begin{figure}
\centering
\centerline{\psfig{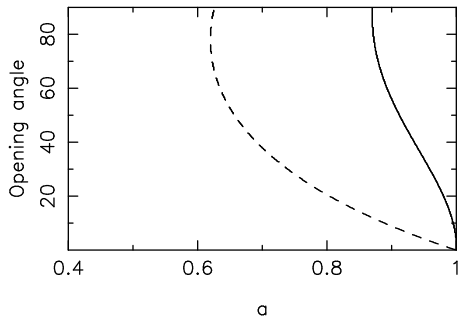}}
\caption{This shows how $a$, as defined by equation~\ref{equ_par_d2}
  is limited by our data as a function of
  the opening angle of QSOs (solid line). The parameter space above
  and to the right of the
  line is inconsistent with our results. For comparison with Fine~et
  al.~(2008) we also constrain $a$ by their parametrisation for the
  BLR (dashed line)}
\label{fig_oang_a_c1}
\end{figure}

Fig.~\ref{fig_oang_a_c1} shows how the limit on $a$ depends on the
assumed opening angle to QSOs (solid line). To facilitate comparisons
with Fig.~15
in \citealt{me2} we have also plotted the same constraint assuming the
parametrisation used in that paper (dashed line).
In both cases very disk-like BLRs
are ruled out at all assumed opening angles. This is an extension of
the argument that, if the BLR is a confined disk, we should find more
narrow-line QSOs. Since we do not, the BLR must have a significant
non-disk component \citep{ost77}.


\begin{figure}
\centering
\centerline{\psfig{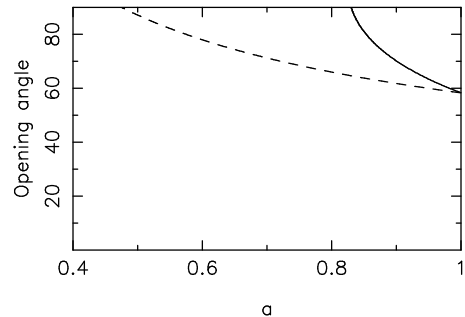}}
\caption{This shows how $a$, as define by equation~\ref{equ_par_w2}
(solid line) is limited by our data as a function of opening
  angle. Again we also show the constraint following Fine~et
  al.~(2008) for comparison (dashed line).}
\label{fig_oang_a_c2}
\end{figure}

We can perform the same tests assuming a velocity field that is
constrained in the polar direction, potentially a BLR which is part of
a wind. In this case our parameterisation for the BLR becomes:
\begin{equation}
\label{equ_par_w2}
{\rm line\,width}\propto \sqrt{a^2\cos^2(\theta) + (1-a^2)}.
\end{equation}
Fig~\ref{fig_oang_a_c2} plots the results for this
model. Here, even with very constrained BLRs (i.e. $a\sim1$), we can
only rule out the model if the opening angle to QSOs is large
($>60^\circ$).

\section{Conclusions}

We have performed an extensive analysis of the \civ\ line width
distribution in QSO spectra from the 2SLAQ, 2QZ and SDSS.
We reviewed the three most common methods for fitting \civ\ emission in
QSO spectra and performed a detailed
comparison between these, using both composite spectra and individual
fits to our whole
dataset. Based on these results we have chosen the procedure that we
believe is the most physically motivated and least biased to employ
for our analysis. Furthermore we have developed a procedure for
identifying absorption features in spectra and removing BAL systems from
the dataset.

Applying our routine to spectra from the SDSS, 2QZ and 2SLAQ surveys
we have measured the \civ\ line width distribution for QSOs in the
redshift range $1.5 < z < 3.3$ spanning a magnitude range of
$-24 > M_r(z=2.5) > -29$. We find that the line width
vs. luminosity plot (Fig.~\ref{fig_M_lw_c}) shows many similarities
with the equivalent plot for \mgii. However, this appears to be where
the similarity ends.

We compare the \civ\ and \mgii\ line widths calculated in spectra
that have both lines, and find considerable scatter in the
comparison. We find that the dispersion in \civ\ line width is smaller
than that for \mgii\ and does not show the same trend with luminosity
that \mgii\ exhibits.

These results are discussed in terms of virial SMBH mass estimation. We
show that, for both \mgii\ and \civ\ estimators, the line width term
contributes considerably less dynamic range to the resulting mass
estimate than the quoted error on the estimate. The dynamic range
found in virial SMBH mass estimates comes (almost entirely) from their
luminosity term, and this is solely responsible for the consistency of
virial masses based on the two lines since the line widths do not correlate.

Finally the results are discussed in terms of BLR dynamics. We show
that given the small scatter in \civ\ line widths the \civ\ BLR cannot
be a flat disk. We parametrise models for a hybrid BLRs which include
a random/isotropic velocity component and show how our results can be
used to constrain these models.

\section{Acknowledgements}

We would like to thank all our colleagues who gave useful input into
this work. In particular we would like to thank Gordon Richards for
help with proofing the paper. In addition we thank to all of the good people
at the University of Sydney for their help, their advice and their
support.

SMC acknowledges the support of an Australian Research Council QEII
Fellowship and a J G Russell Award from the Australian Academy of
Science.
JBH is supported by a Federation Fellowship from the Australian
Research Council.

The authors would like to thank all the present and former staff of the
Anglo-Australian Observatory for their work in building and operating
the 2dF facility.  The 2QZ and 2SLAQ are based on
observations made with the Anglo-Australian Telescope and the UK
Schmidt Telescope as well as the Sloan telescope.

In addition the authors would like to thank the SDSS project from
which much of the data in this paper was obtained.
Funding for the SDSS and SDSS-II has been provided by the Alfred
P. Sloan Foundation, the Participating Institutions, the National
Science Foundation, the U.S. Department of Energy, the National
Aeronautics and Space Administration, the Japanese Monbukagakusho, the
Max Planck Society, and the Higher Education Funding Council for
England. The SDSS Web Site is http://www.sdss.org/.

The SDSS is managed by the Astrophysical Research Consortium for the
Participating Institutions. The Participating Institutions are the
American Museum of Natural History, Astrophysical Institute Potsdam,
University of Basel, University of Cambridge, Case Western Reserve
University, University of Chicago, Drexel University, Fermilab, the
Institute for Advanced Study, the Japan Participation Group, Johns
Hopkins University, the Joint Institute for Nuclear Astrophysics, the
Kavli Institute for Particle Astrophysics and Cosmology, the Korean
Scientist Group, the Chinese Academy of Sciences (LAMOST), Los Alamos
National Laboratory, the Max-Planck-Institute for Astronomy (MPIA),
the Max-Planck-Institute for Astrophysics (MPA), New Mexico State
University, Ohio State University, University of Pittsburgh,
University of Portsmouth, Princeton University, the United States
Naval Observatory, and the University of Washington. 

\bibliographystyle{mn2e}
\bibliography{bib}

\appendix

\section{Fitting the \civ\ line} \label{sec:fitting}

QSO emission in the vicinity of the \civ\ line is complex and
there is no simple, accepted prescription for fitting its profile.
Complex iron emission is likely to pervade the \civ\ line region. In
addition, the red wing of the line is contaminated by the weak \heii\
$\lambda$1640 and \oiii\ $\lambda$1663 lines (the \oiii\ line may also be
blended with Al\,{\sc ii} $\lambda$1671, but we shall just refer to this
blend as \oiii), as well as a significant
contribution from an unidentified source at $\sim1600$\,\AA\
(e.g. \citealt{wil84,boy90,lao94,van01}). Much of the difficulty in
fitting the \civ\ line is due to this 
unidentified emission in the red wing of the line.

Furthermore, high S/N spectra have shown evidence for \niv\
$\lambda$1486 and \siii\ $\lambda$1531 at low levels in the blue wing
of the line (e.g. \citealt{c+v90,boy90,lao94,v+w01}).

Most of these contaminating lines are weak compared
with \civ\ and can be corrected for. The primary difficulty when
fitting the \civ\ region of quasar spectra is
the excess emission at $\sim1600$\,\AA. Conflicting fitting
prescriptions that correct for this feature lead to systematic biases
when parameterising the \civ\ line.

\subsection{The $\sim1600$\,\AA\ feature}

\begin{figure}
\centering
\centerline{\psfig{file=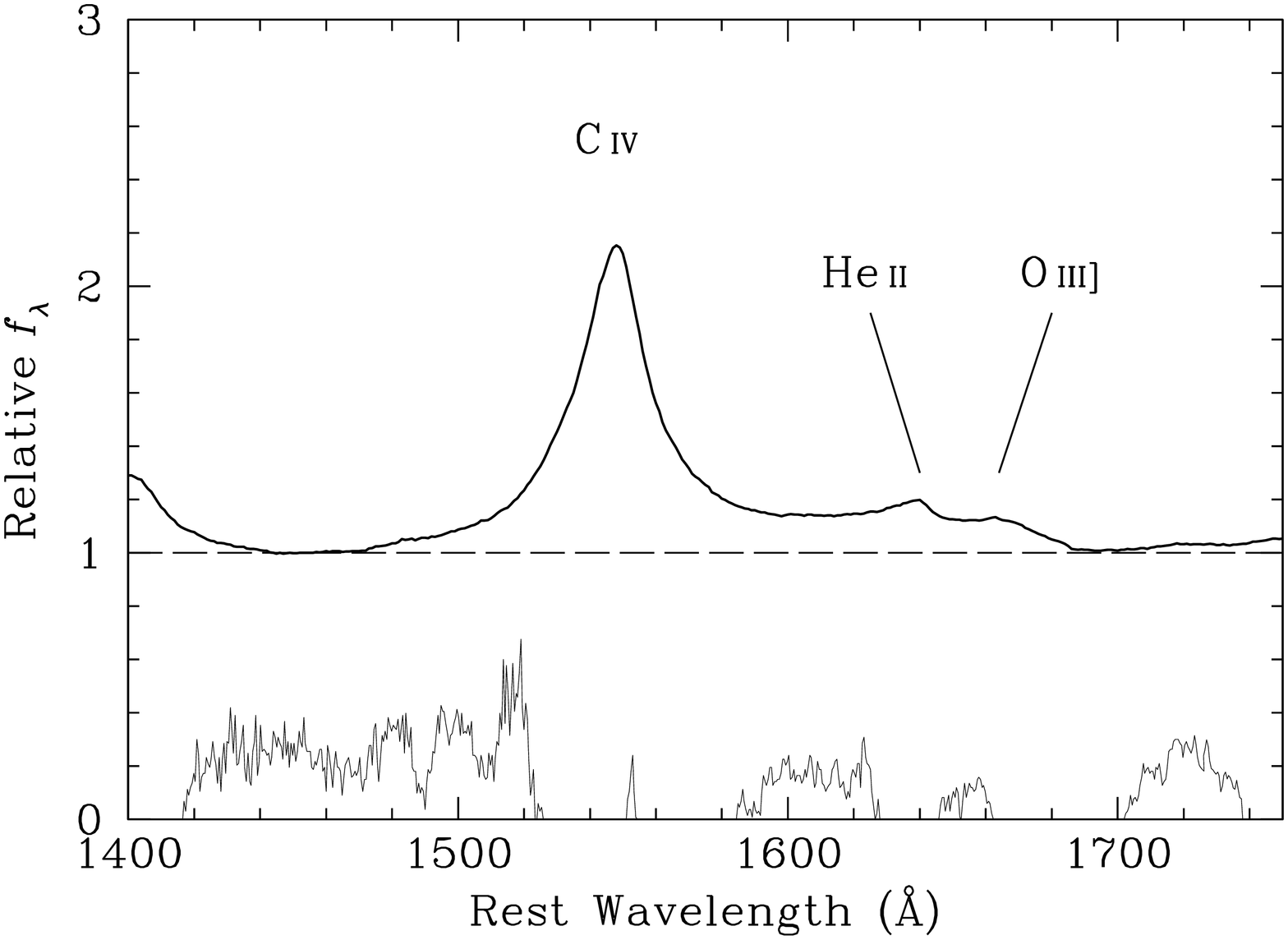,width=9.0cm,angle=-90}}
\caption{The \civ\ region of the 2QZ QSO composite spectrum (heavy
line) with known emission features labelled. In
constructing this composite each spectrum is normalised to a fitted
continuum. The normalised continuum is shown as a dashed line 
to highlight excess emission. The fine line at the bottom of the
shows the Vestergaard \&\ Wilkes (2001) template for iron emission.
We have scaled the iron template by an arbitrary constant for ease of
comparison in the figure.}
\label{fig_civ_fe}
\end{figure}

Fig.~\ref{fig_civ_fe} shows the \civ\ region from the 2QZ
QSO composite \citep{croom02}. Obvious emission features due to
\civ, \heii\ and \oiii\ are labelled. The composite shows no
evidence for \niv\ or \siii\ in the blue wing. However, there is clearly
emission around 1600\,\AA\ that can not easily be described as
a combination of the labelled lines.

\citet{wnw80} suggested there should be significant \feii\ emission in
the $\sim$$1610-1680$\,\AA\ range based on models for \feii\ emission due
to collisional excitation. More recent models for quasar iron
emission also show significant flux in this region
\citep{s+p03}. However, the models do not contain enough flux at
$1600-1610$\,\AA\ to accurately fit observed quasar spectra.

The empirical iron template of \citet{v+w01} shows emission between
1590 and 1620\,\AA, but again not at the levels required to accurately fit
observed spectra. Fig.~\ref{fig_civ_fe} shows the \citet{v+w01} iron
template. Comparing the template with the composite spectrum at 1450 or
1720\,\AA\ and 1600\,\AA\ shows why the excess emission at 1600\,\AA\
cannot be accurately described with this template.

The lack of of flux in the vicinity of \heii\ and \oiii\ in the
Vestergaard \& Wilkes iron template is a result of their removal of
Gaussian fits to the \heii\ and \oiii\ lines. It may be that in
doing so the iron template has been over corrected and there may, in
fact, be residual iron emission in these regions.

\subsection{Techniques for fitting the \civ\ line}

Due to uncertainty in identifying the source of the emission
around \civ, there is no standard prescription for
analysing the \civ\ line, and previous studies have implemented various
techniques for fitting the \civ\ region. These approaches can all be
classified as three different ways of dealing with the emission at 1600\,\AA.

\begin{description}
\item[1)]{Assume the emission at 1600\,\AA\ is a red wing of \civ.}
\item[2)]{Assume the emission at 1600\,\AA\ is due to another species and
try to correct for it.}
\item[3)]{Try to fit around the 1600\,\AA\ feature without attempting to
explain it.}
\end{description}

We examine each of these line fitting proceedures in turn.

\subsubsection{The 1600\,\AA\ feature as a red wing to the \civ\ line}

\begin{figure*}
\centering
\centerline{\psfig{file=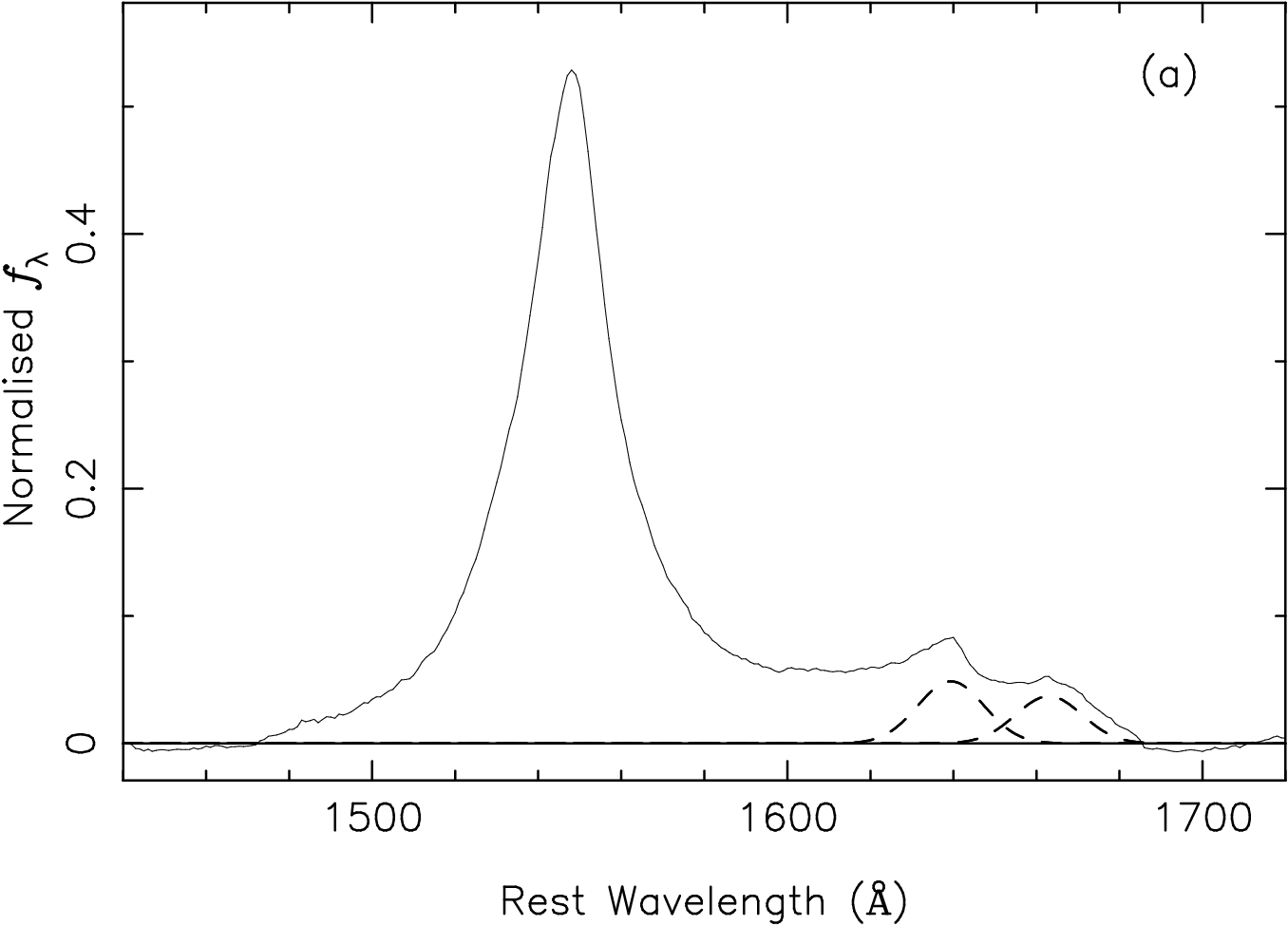,width=7.5cm,angle=-90}\hspace{0.5cm}\psfig{file=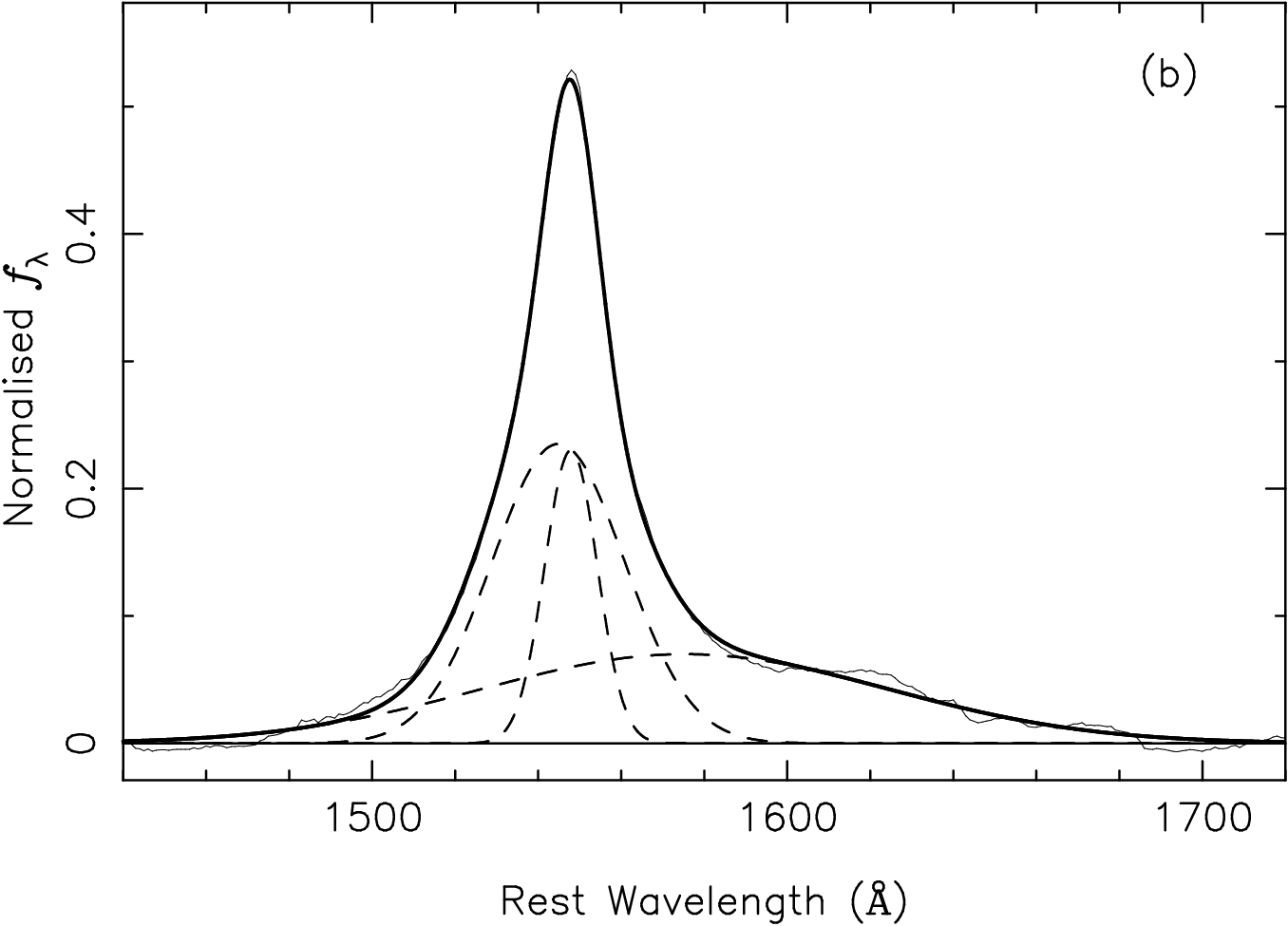,width=7.5cm,angle=-90}}
\caption{Fitting the \civ\ region of the 2QZ composite by assigning
three Gaussian
components to the \civ\ line, and one each for \heii\ and
\oiii. \empha\ The continuum-subtracted composite with dashed lines
showing the fits to \heii\ and \oiii. \emphb\ The composite after
subtracting the Gaussians fit to \heii\ and \oiii. Dashed lines
show the three components of the \civ\ line and the heavy lines shows
their sum.}
\label{fig_comp_fit_3g}
\end{figure*}

One of the more common fitting proceedures treats the excess
emission at 1600\,\AA\ as an extended red wing of the \civ\ line
\citep{lao94,me1,shen08}. A continuum is fitted to the
local region ($\sim1450-1700$\,\AA) with or
without an iron template included. Then single Gaussians are fitted to
contaminating features (\heii, \oiii, \niv\ etc) along
with three Gaussians to describe the \civ\
line. Fig.~\ref{fig_comp_fit_3g} illustrates this procedure
applied to the 2QZ QSO composite.

Three Gaussians are needed to fit the \civ\ line since the residual
after the continuum and other lines have been subtracted
is markedly asymmetric. However,
Fig.~\ref{fig_comp_fit_3g}\emphb\ shows that the resulting fit to the
\civ\ region is able to accurately model the spectrum. The flaw in
the prescription is that it assigns
a significant amount of \civ\ emission at implausible velocity shifts
from the line centre \citep{lao94}, and that it imposes a strong
asymmetry on the \civ\ line which is not obvious in the
original spectrum.

\subsubsection{The 1600\,\AA\ feature as \heii\ emission}

\begin{figure*}
\centering
\centerline{\psfig{file=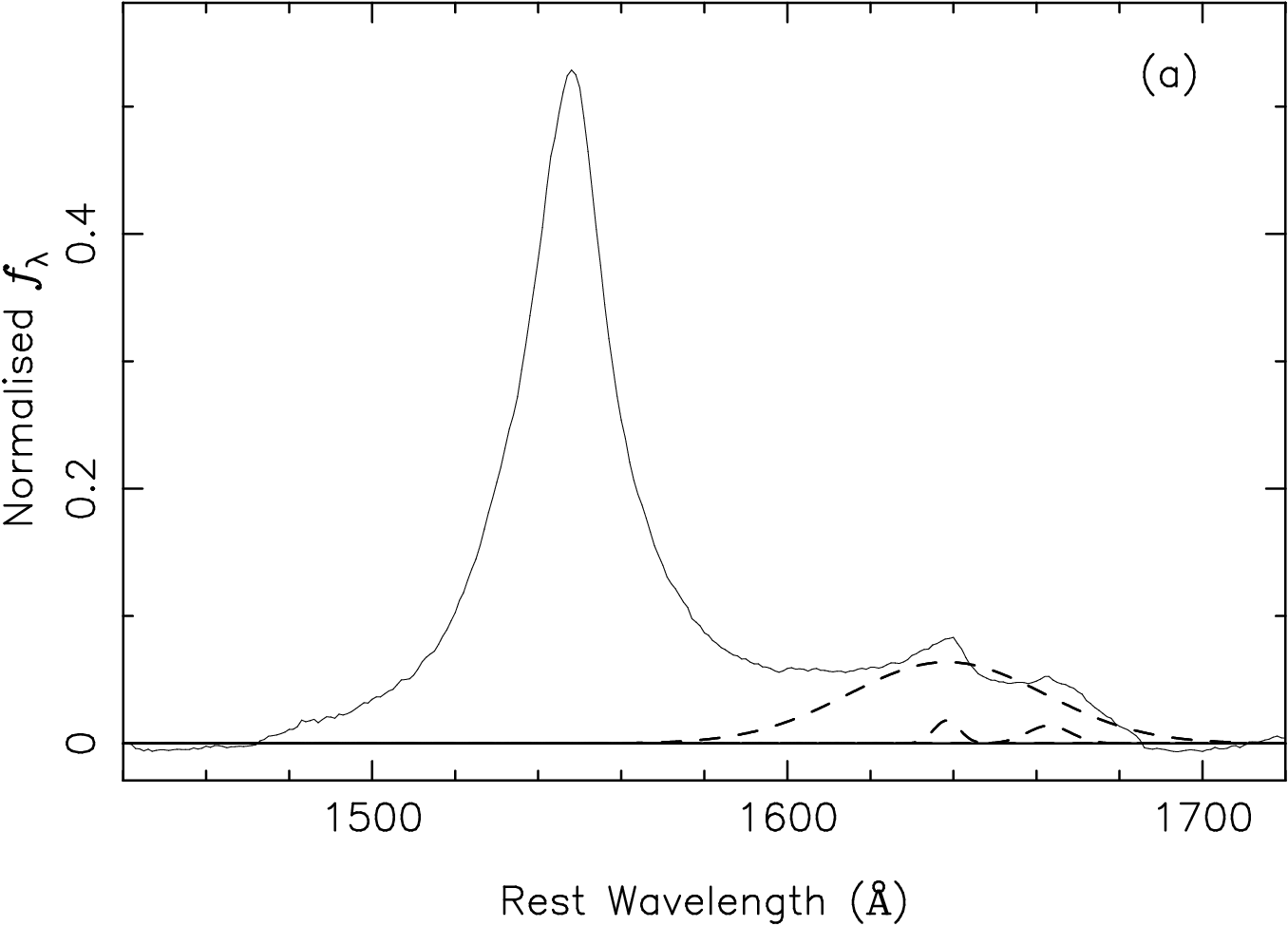,width=7.5cm,angle=-90}\hspace{0.5cm}\psfig{file=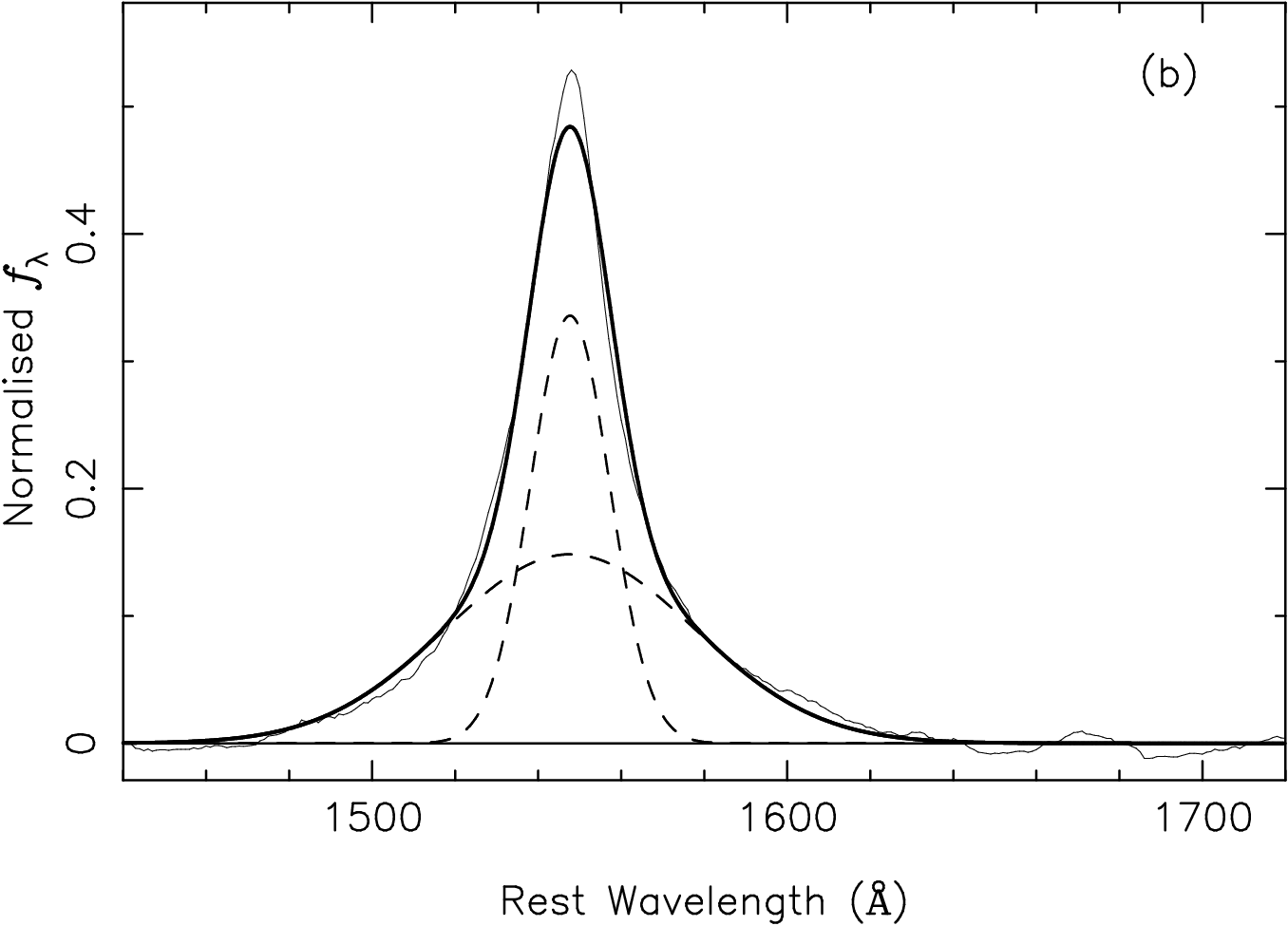,width=7.5cm,angle=-90}}
\caption{Fitting the \civ\ region by assigning both a broad and narrow
component to the \heii\ line. Here the observed emission
redwards of 1600\,\AA\ is removed as \heii\ and the result is a
relatively symmetric profile
for \civ. In \empha\ the continuum subtracted 2QZ QSO composite is
shown along with the three Gaussians fitted to the \heii\ (two) and
\oiii\ (one) features. In \emphb\ these lines have been subtracted from the
composite and two Gaussians have been fitted to the \civ\ line; their sum
is shown as the heavy line. Note that all Gaussian are fit
simultaneously in the procedure.}
\label{fig_comp_fit_2he}
\end{figure*}

\citet{croom02} fit the emission redwards of \civ\ with a single
broad Gaussian centred at 1640\,\AA\ associated with \heii\
emission. This effectively removes the strong emission feature at
$1600-1700$\,\AA; however, it does not accurately remove residual
features in the region. \citet{shang07} go a step further and fit
for narrow \oiii\ and \heii\ emission on top of broad \heii\
emission. These additional components allow one to accurately
reproduce the shape of quasar spectra in the \civ\ region; an example fit
is given in Fig.~\ref{fig_comp_fit_2he}.

Fig.~\ref{fig_comp_fit_2he}\empha\
shows the continuum subtracted 2QZ composite spectrum along with
dashed lines showing the two Gaussians fitted to \heii\ and the one to
\oiii. Fig.~\ref{fig_comp_fit_2he}\emphb\ shows the residual spectrum
once these components have been subtracted. Comparing
Fig.~\ref{fig_comp_fit_2he}\emphb\ with
Fig.~\ref{fig_comp_fit_3g}\emphb, we find the strong asymmetry that
results from assuming the 1600\,\AA\ feature is \civ\
emission is not evident when applying this prescription for fitting. Since
the resulting \civ\ line is almost symmetric it can be relatively well
modelled by a sum of two Gaussians as in
Fig.~\ref{fig_comp_fit_2he}\emphb.

\begin{figure}
\centering
\centerline{\psfig{file=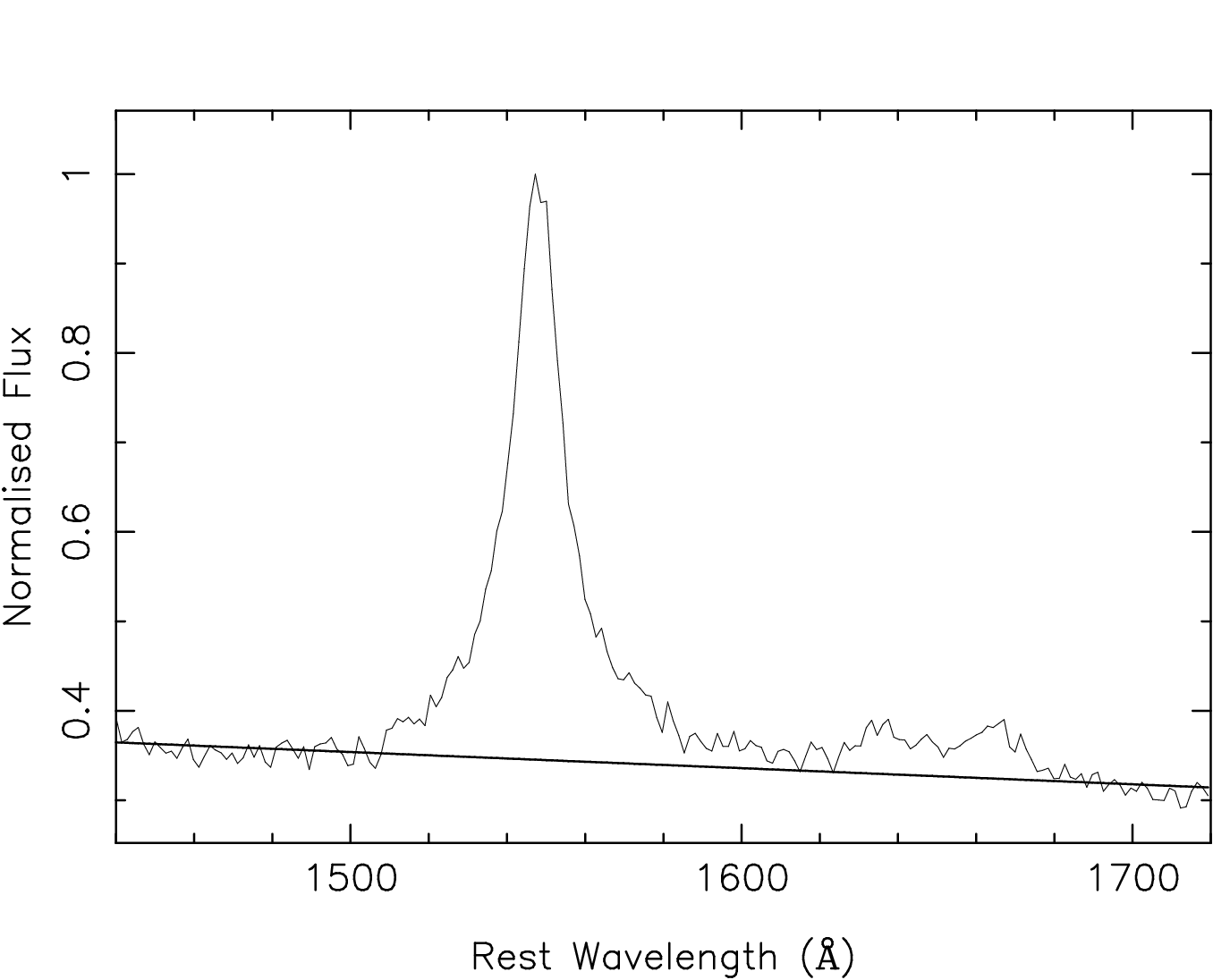,width=7.0cm,angle=-90}}
\caption{The \civ\ region from the spectrum of QSO
J222203.6-320437 taken during the 2QZ survey. This spectrum shows
\heii\ and \oiii\ features, but
no other significant emission redwards of $\sim1600$\,\AA.}
\label{fig_civ_no_wing}
\end{figure}

There are two main strengths to this second fitting prescription. Firstly,
each separate component of the fit is assigned to 
a particular ion; hence it can be argued that the prescription makes sense
physically. Secondly, the fit preserves the symmetry of the \civ\
line. While we do not know that the \civ\ line is intrinsically
symmetric, it is difficult imagine that emission extending to $>20$,000\,km/s
is associated with the \civ\ line. Furthermore, many individual high
S/N spectra show strong \civ\ emission but very little emission in the
red wing (e.g. Fig.~\ref{fig_civ_no_wing}).

We must remember that just because we have fit the 1600\,\AA\ feature with a
broad \heii\ component does not necessarily mean that \heii\ is
responsible for the emission. At the point where the broad \heii\
component becomes blended
with the red wing of \civ\ it is not clear that extrapolating the
Gaussian fits will remove contaminating emission from the \civ\ line
correctly.
Hence the main concern with this fitting proceedure derives
from its strength: because we are assigning the 1600\,\AA\ emission to
\heii\ the fit appears to be physical, and this clouds the fact that we may be
introducing an unknown systematic into the results.

\subsubsection{Fitting around the 1600\,\AA\ feature}

A third option for fitting the \civ\ line is to accept that it is
unclear how to correct for the emission redwards of \civ\ and to try
and perform the simplest, non-parametric correction as
possible. \citet{wil08} fit a linear continuum
between small windows centred at 1480 and 1690\,\AA, and then
calculate central moments over the interval $1496-1596$\,\AA\ to
describe the \civ\ line.

This approach has the advantage of simplicity and does not rely on
imposing a specific profile (Gaussians in the examples discussed
above) on spectral features. Since no attempt is made to correct for the
excess flux at 1600\,\AA, this prescription will likely have larger
systematic errors than fitting for \heii\ emission in this
region. However, the systematics are also likely to be more consistent
when compared to other fitting proceedures since there are many fewer
fitted parameters.

\begin{figure*}
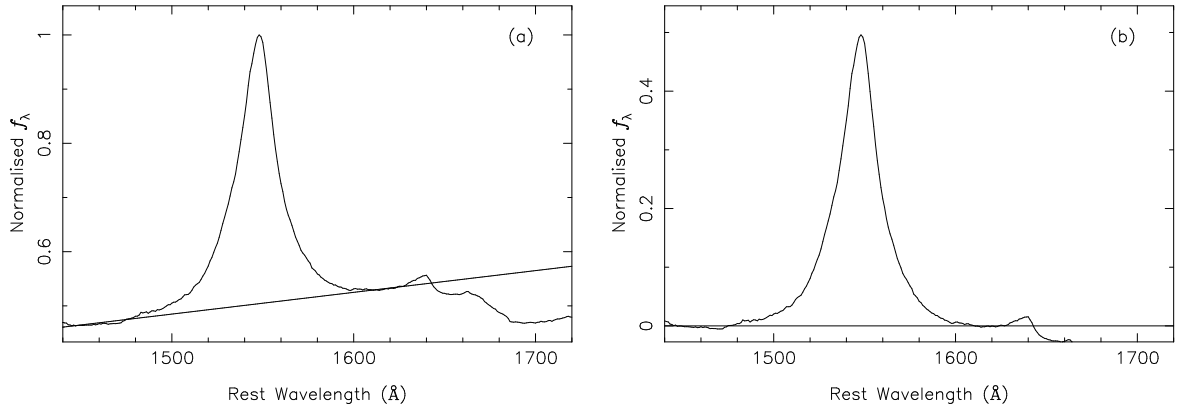

\centering
\centerline{\psfig{file=civ_comp_fit_lcont_1.ps,width=7.5cm,angle=-90}\hspace{0.5cm}\psfig{file=civ_comp_fit_lcont_2.ps,width=7.5cm,angle=-90}}
\caption{The simplest prescription for fitting the \civ\ line. \empha\ 
shows the 2QZ QSO composite with a linear fit between windows at 1450
and 1610\,\AA. In \emphb\ the continuum has been subtracted from the spectrum.}
\label{fig_comp_fit_lcont}
\end{figure*}

As a small adjustment to \citet{wil08} we also consider a
prescription
where the continuum is fitted between 1450 and 1610\,\AA, and then
calculate parameters for \civ\ in this region as shown in
Fig.~\ref{fig_comp_fit_lcont}. The limits are chosen as the local
minima either side of the \civ\ line, and can be thought of as the
regions where \civ\ emission ceases to dominate the spectral
shape.

The fitting proceedure illustrated in Fig.~\ref{fig_comp_fit_lcont} has
the advantages of that used by \citet{wil08}, while
also making a simple correction for emission in the red wing of the
line. The primary problems associated with this fit are the predefined
\civ\ line region and the highly artificial way we have corrected for
contaminating emission. Each of these issues will systematically bias
our derived line widths. This procedure can be considered as an
extreme limit to plausible fits.

In the next section we discuss calculations of the width of
the \civ\ line, before making a quantitative comparison between the three
fitting proceedures in section~\ref{fm_comp}.

\subsection{Calculating the line width}

There are several techniques for parameterising the width of a spectral
line. Three commonly used measures are the full width at half maximum
(FWHM), the inter-percentile velocity (IPV) width, and the dispersion
(or second moment) of a line.

The FWHM is most commonly employed as the line width parameter when
considering QSO broad lines. It is easy to define and calculate, and
for high S/N spectra gives an accurate line width. However, when
measuring the FWHM directly from low S/N spectra problems arise, both when
defining the maximum flux density of a line and when dealing with
multiple crossings of the half maximum value.

These problems can be circumvented by fitting a model line to the
spectrum and measuring the FWHM of the model rather than from the
spectrum (e.g. \citealt{lao94,me1,shen08}); this of course assumes
that the model gives an accurate representation of the line.

In the following analysis we measure the FWHM from our model fits to
the spectra. Our models are composed of several Gaussian
components. These Gaussians are fit to our data with the {\sc mrqmin}
routine \citep{press92} which returns the fitted parameters along with
their covariance matrix. The error on the FWHM is calculated
incorporating the covariance between the fitted parameters. However,
since {\sc mrqmin} does not take a covariance matrix as input,
covariance in the continuum-subtracted spectrum (see
e.g. \citealt{cardiel98}) is not incorporated in 
the final error estimate. Hence the error on the FWHM we calculate will
be underestimated.

Another measure of the line width that is becoming increasingly
widespread is the dispersion or second moment of the line
\citep{fr+me00,v+p06,wil07}. However, we find that the excessive weighting this
measure assigns to the values of pixels in the wings of the lines makes
it an unreliable estimator of line width in low S/N spectra.

Inter-percentile velocity (IPV) widths offer a third parametrisation
of the line widths (e.g. \citealt{whit85,me2}). While at first
glance the process of measuring an IPV width is similar to measuring
the FWHM, the dependence of IPV widths on the cumulative flux
distribution rather than the flux density at a given point makes the
IPV measurements considerably more robust with respect to noise in the
spectrum. The IPV width can be very susceptible to uncertainty in the
continuum placement. However, even in low S/N spectra a linear continuum can
be fit to a relatively high degree of accuracy and precision given a
modest spectral region to fit over.

Like the dispersion, IPV widths are somewhat affected by noise in the
wings of lines; in particular this can affect the total flux of a
line and how one defines the zero-point of the cumulative flux
distribution. However, when calculating the dispersion the weight
given to a single pixel is proportional to the square of the
displacement of that pixel from the line centre. This power-of-two
dependence makes the dispersion highly susceptible to noise in the wings of a
line; this is not a problem for IPV widths.

In this analysis we calculate the 50\.\%\ IPV width for the \civ\
lines in our sample (i.e. the width between the 25\,\%\ and
75\,\%\ crossings of the cumulative flux distribution). We calculate
the IPV width directly from the spectrum, interpolating between
pixels either side of the crossings. Errors on the IPV widths are
calculated from the spectral variance 
array including the contribution of covariance introduced by the iron
and continuum subtraction.

Finally, for any line width measure, we subtract the resolution of the
spectrograph in quadrature from the measured line width under the
assumption of a Gaussian profile for both the emission line and
instrumental resolution.

\subsection{Comparisons between fitting proceedures} \label{fm_comp}

In this section we compare the above prescriptions for fitting the \civ\
line to highlight the possible biases introduced by each.
The precise proceedures implemented in each case are:

\begin{description}
\item[1)]{
We fit a linear continuum under the \civ\ region between two 45\,\AA\
wide spectral windows at $1430<\lambda<1475$\,\AA\ and
$1680<\lambda<1725$\,\AA. The  continuum is subtracted from
the spectrum and we then fit five Gaussians to the residual. Two of
these have their wavelengths fixed at the expected wavelength of
\heii\ and \oiii; the final three are taken to describe the \civ\
line. 

Both the FWHM and 50\,\%\ IPV widths are measured for the line. The
FWHM is measured from the three-Gaussian model for the line while the
IPV width is calculated directly from the spectrum.
}
\item[2)]{
We perform the same continuum fit as in (1). Five Gaussians
are fitted to the continuum-subtracted spectrum, two are fixed to
\heii\ and one to \oiii\ 
while the final two Gaussians describes the \civ\ line. In the fit, the
two Gaussians which describe \heii\ and the two which describe \civ\
have their central wavelengths fixed to the same value. The three
Gaussians which were fitted to \heii\ and \oiii\ are then subtracted
from the spectrum;
the FWHM is calculated from the double-Gaussian model for the line.
}
\item[3)]{
As a final prescription we fit a linear continuum between 20\,\AA\ wide
windows centred at
1450 and 1610\,\AA. The continuum is subtracted and two Gaussians are fit to
the residual \civ\ line with their central wavelengths tied together.
}
\end{description}

We compare these prescription in two ways. Firstly, we compare fits to the
high S/N 2QZ QSO composite spectrum. Secondly, we apply each of these
routines to our dataset and compare the overall results.

\subsubsection{Detailed fits to the 2QZ composite}

\begin{figure}
\centerline{\psfig{file=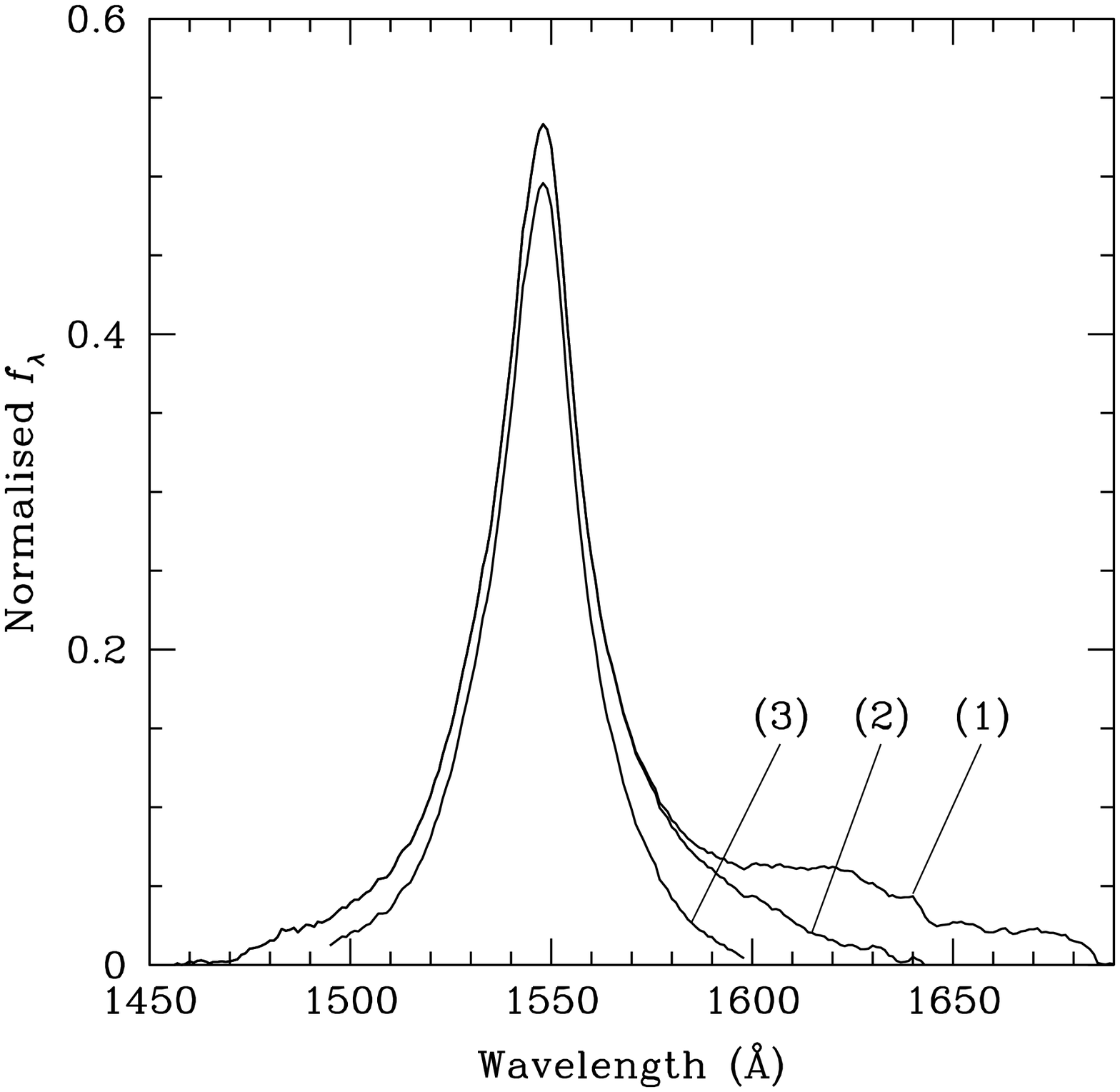,width=8.5cm}}
\caption{Comparison of the residual \civ\ line when fitted using the three
prescriptions described in the text. (1) assumes the \civ\ line is composed
of three Gaussians and fits single Gaussians to \heii\ and \oiii. (2)
fits both broad and narrow components to \heii\ and (3) takes a simple
linear continuum fit between predefined spectral windows either side
of the \civ\ line.}
\label{fig_civ_comp}
\end{figure}

A quick way to compare the differing fitting proceedures is to compare
the results obtained when fitting the 2QZ QSO composite. The fits to
the composite are shown in
Figs.~\ref{fig_comp_fit_3g}, \ref{fig_comp_fit_2he}
and~\ref{fig_comp_fit_lcont}. Fig.~\ref{fig_civ_comp} shows how
the final \civ\ line profiles differ for each of the three approaches
and table~\ref{tab_civ_comp}
gives a selection of line parameters calculated from the differing fits.

\begin{table}
\centering
\caption{A selection of parameters calculated for the \civ\ line in
the 2QZ composite spectrum
based on the differing fitting proceedures. The flux values have been
normalised to the value given by fit~(3).}
\label{tab_civ_comp}
\begin{tabular}{ccccc}
\hline \hline
Fit & Flux & FWHM & IPV & $2^{\rm nd}$ moment \\
 & (rel. to 3) & km/s & km/s & km/s  \\
\hline
1  & 1.53 & 4600 & 6610 & 8410 \\
2  & 1.32 & 4600 & 4540 & 4800 \\
3  & 1.0  & 4180 & 3390 & 3120 \\
\hline \hline
\end{tabular}
\end{table}

The \civ\ profiles for prescriptionss (1) and (2) are identical except for
the red wing of the line. In addition all profiles are similar in the
core of the line, except fit (3) is somewhat lower since the continuum is
fit higher.

The measured line parameters bear out these differences. The IPV
width and, in particular, the $2^{\rm nd}$ moment of the line are
sensitive to flux in the
wings of the lines. Hence the values for these parameters depend strongly
on the fitting proceedure used. The FWHM does not have a strong dependence
on the line wings 
and is similar for all fits. It is slightly smaller in fit (3)
since we have subtracted off more continuum.

\subsubsection{Overall comparisons when fitting the whole dataset}

\begin{figure*}
\centering
\centerline{\psfig{file=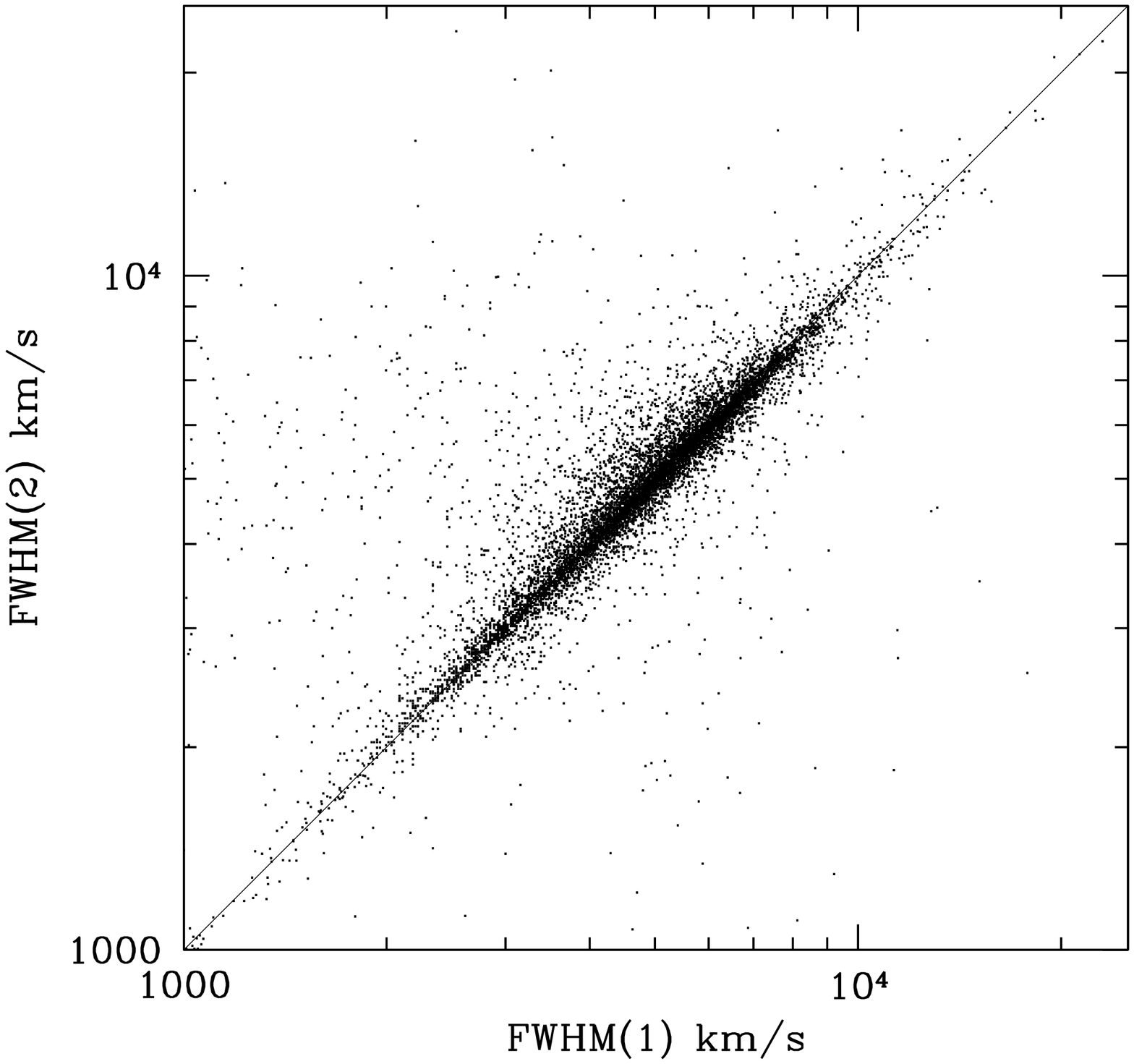,width=6.0cm}\hspace{0cm}\psfig{file=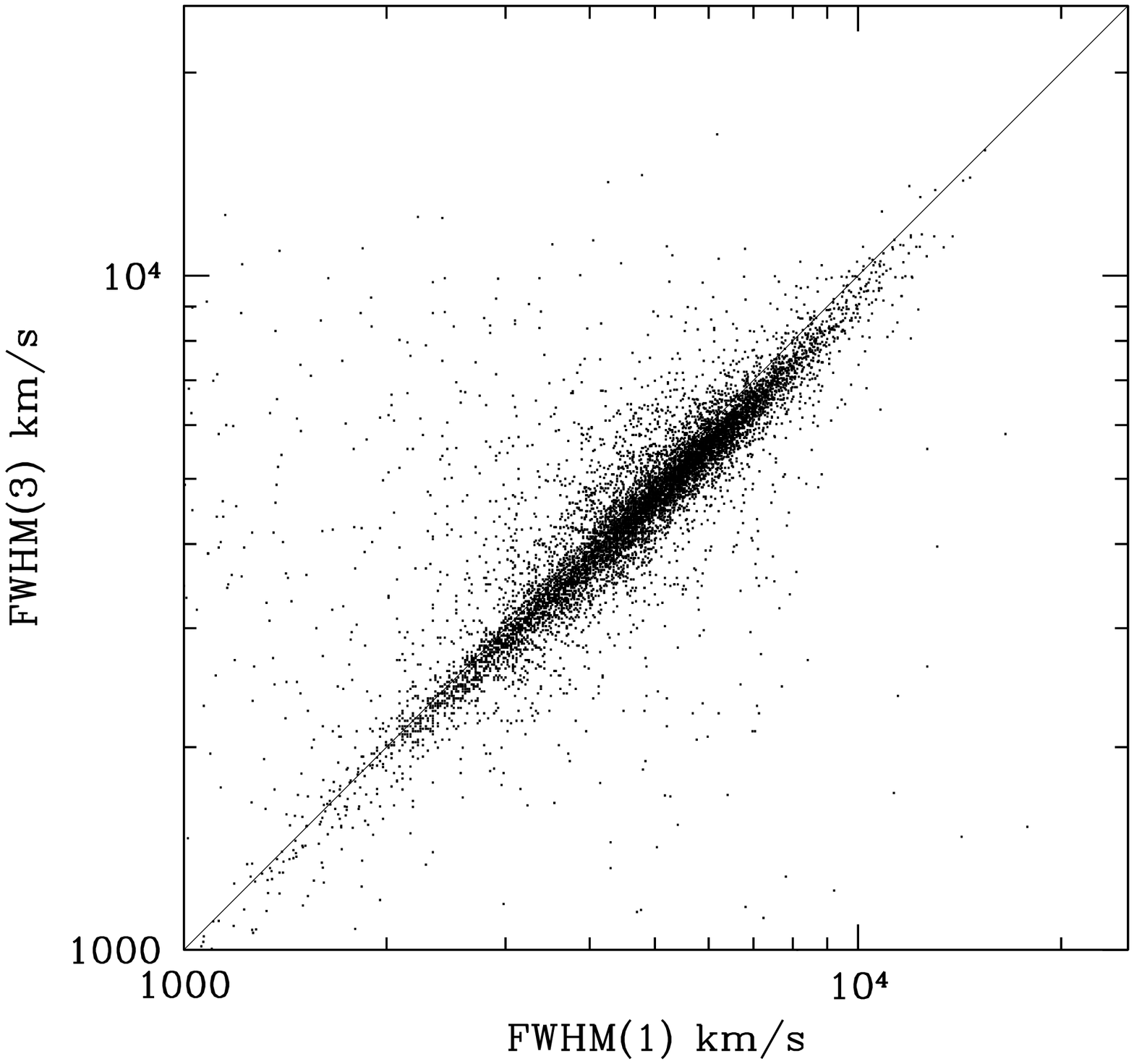,width=6.0cm}\hspace{0cm}\psfig{file=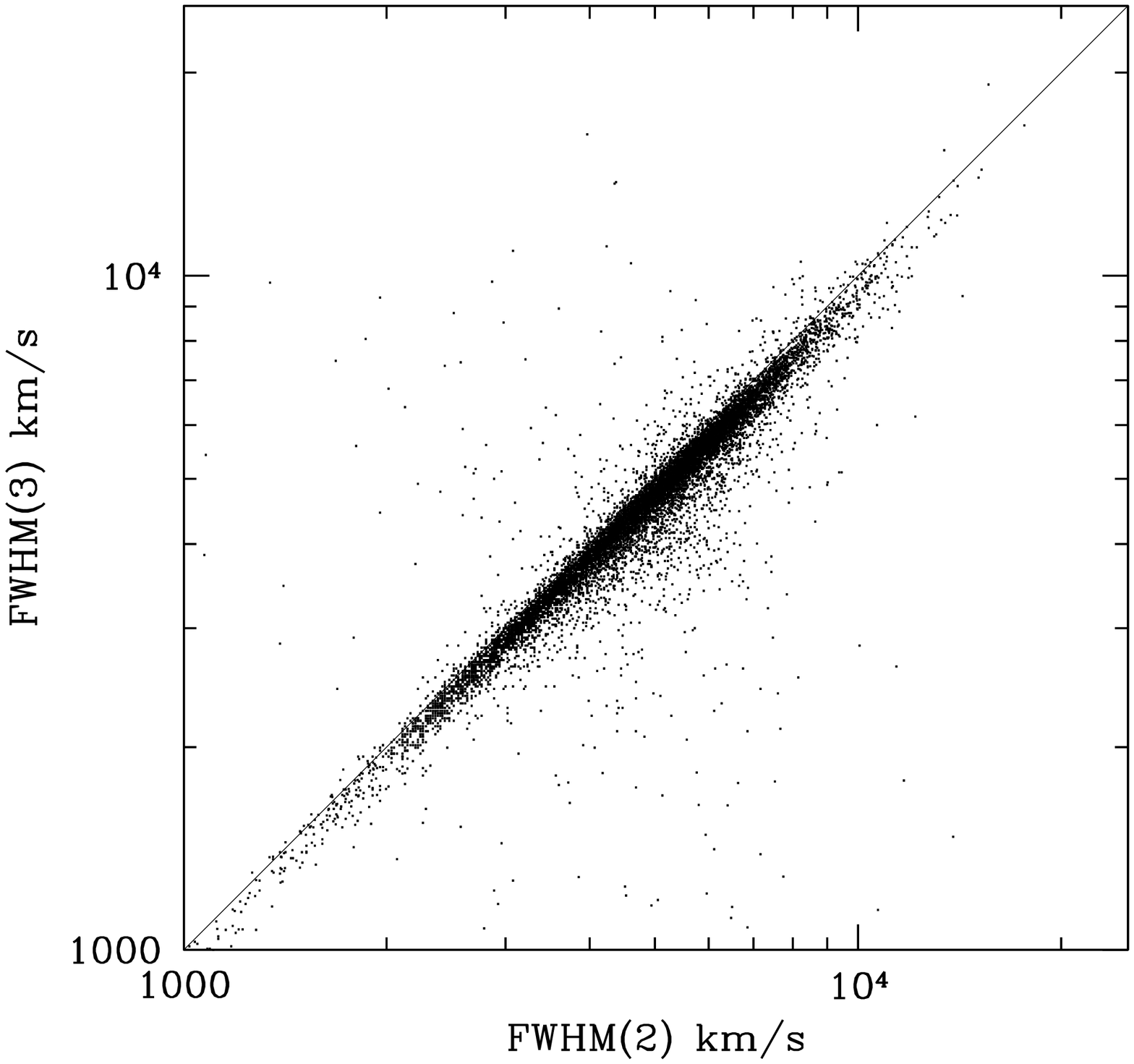,width=6.0cm}}
\vspace{0.3cm}
\centerline{\psfig{file=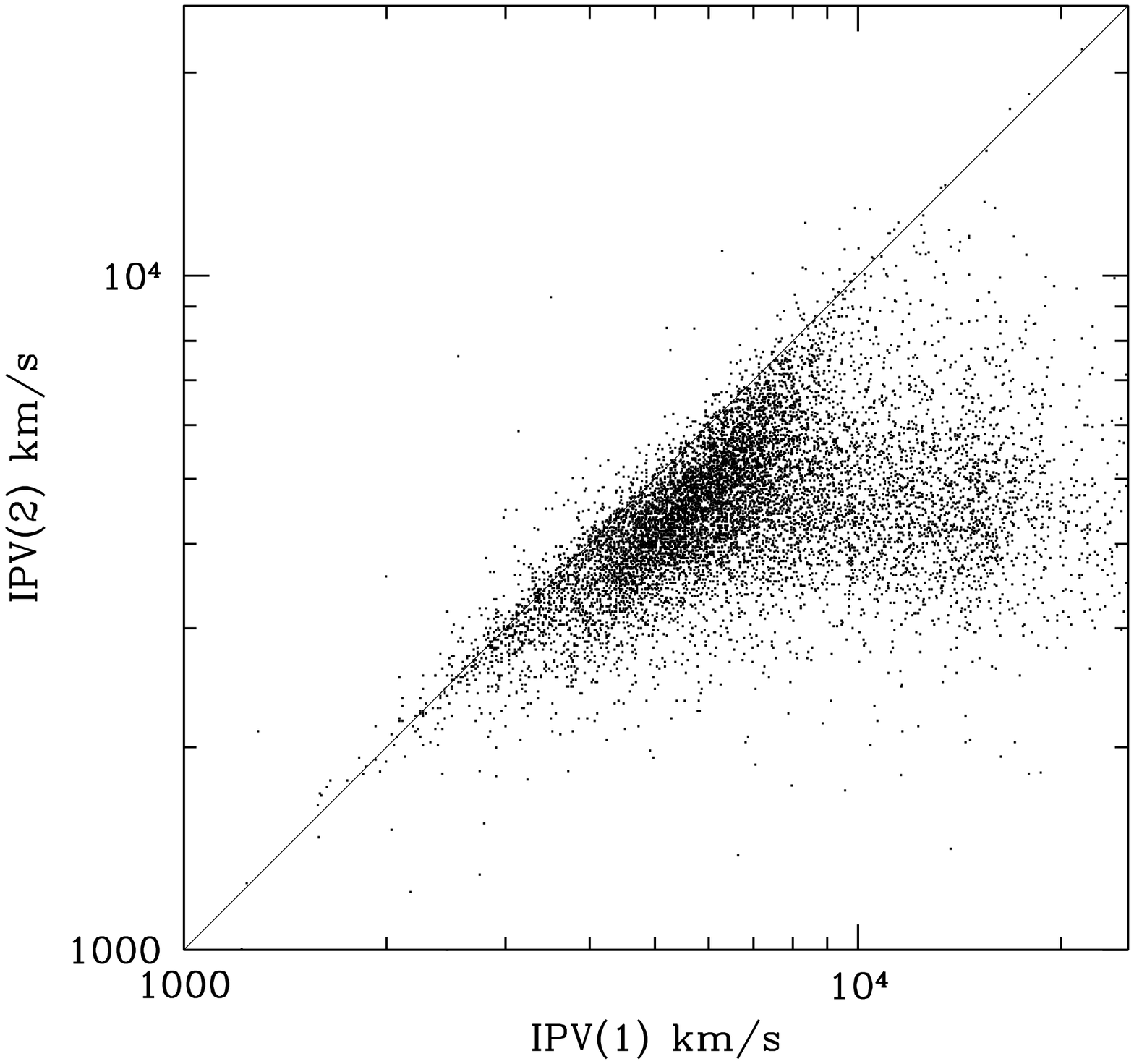,width=6.0cm}\hspace{0cm}\psfig{file=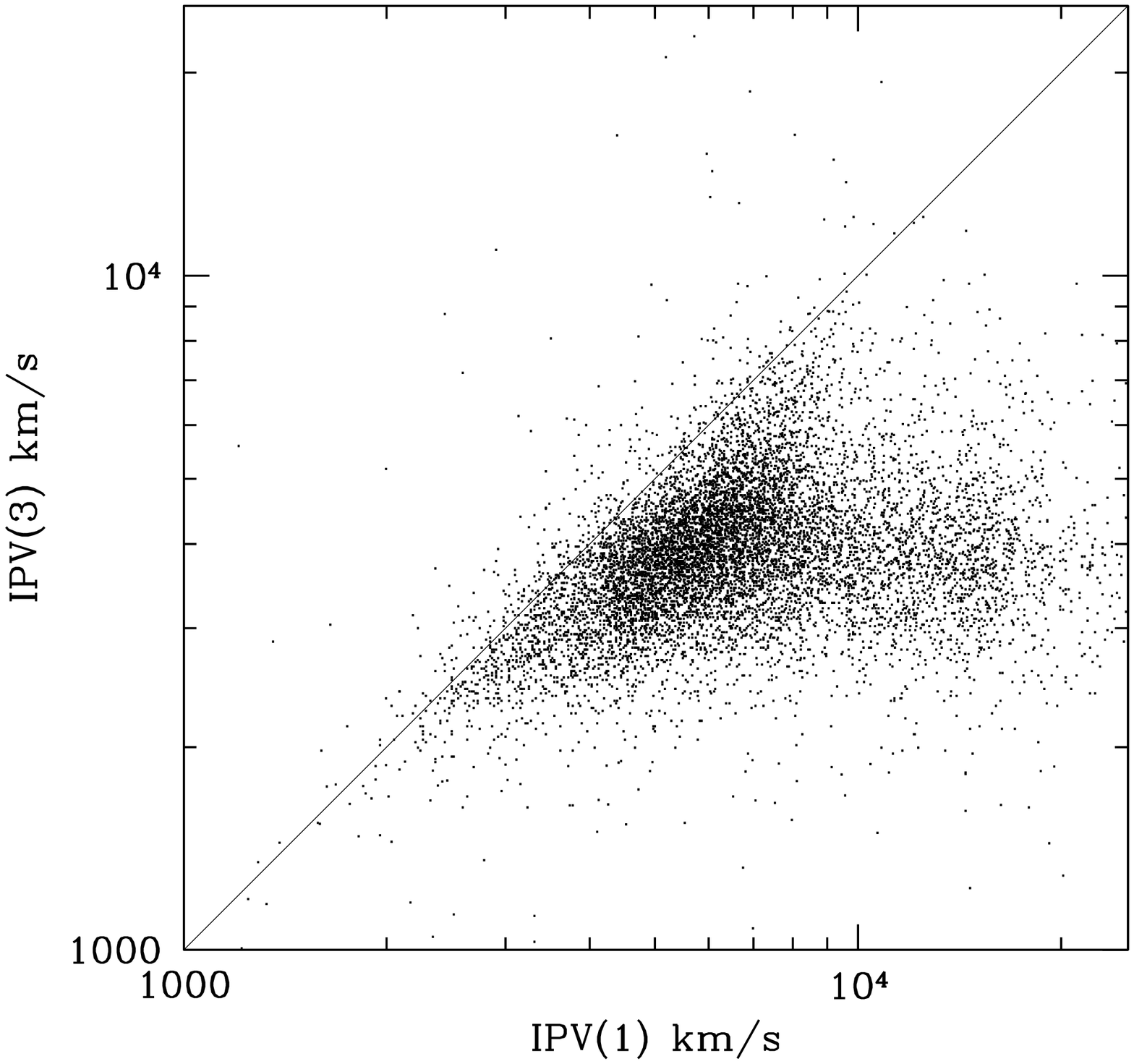,width=6.0cm}\hspace{0cm}\psfig{file=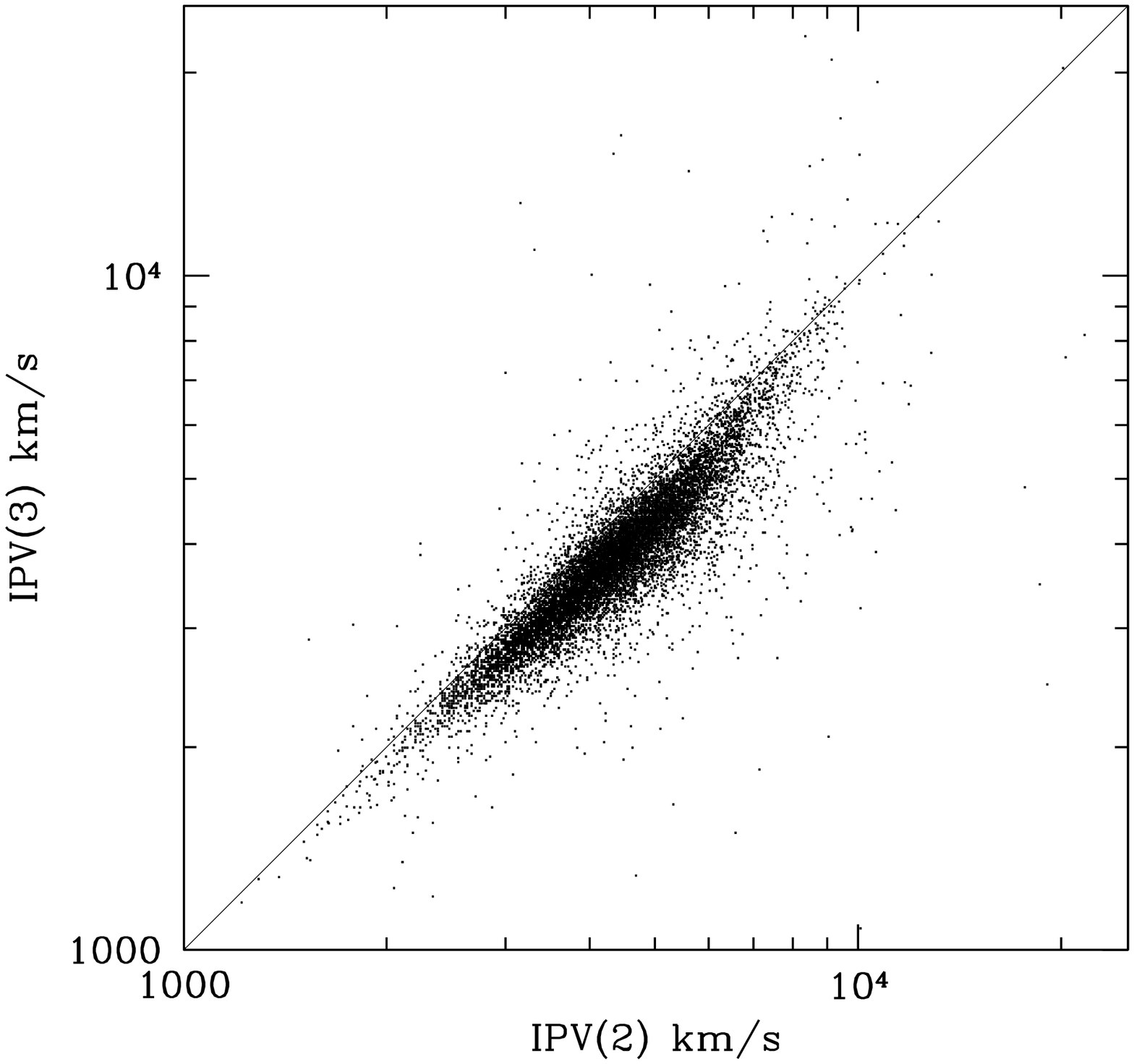,width=6.0cm}}
\caption{Comparisons of the measured line width of the \civ\ line in our whole dataset when applying the three fitting prescriptions described in the text.}
\label{fig_comp_wds}
\end{figure*}

In addition to fitting the high S/N composite we
apply the three fitting proceedures to our entire dataset
to highlight the relative biases of each.
Fig.~\ref{fig_comp_wds} compares the measured FWHM and IPV
width of the \civ\ line when measured via each prescription.

As when fitting the 2QZ composite, each prescription gives equivalent
results for the FWHM of the \civ\ line. Fit~(3) does give slightly
smaller FWHMs by a factor of $\sim1.07$ when compared with the other
two proceedures; this is consistent with the 1.10 ratio obtained when
fitting the composite (table~\ref{tab_civ_comp}). In addition we find
more outliers when comparing the results from (1) with (2) or (3),
suggesting it is a less stable technique.

Prescription (1) gives significantly higher values for the IPV
width when compared with the others. The IPV widths as
measured via (2) and (3) follow a linear relationship. Their
means are offset by a factor of 1.4, comparable to the ratio of 1.5,
between the IPV 
widths measured from the 2QZ composite. A best fit
(found by minimising the 2D $\chi^2$) shows a slight departure from
a linear relation with a gradient of 0.964$\pm$0.002.

\begin{figure*}
\centering
\centerline{\psfig{file=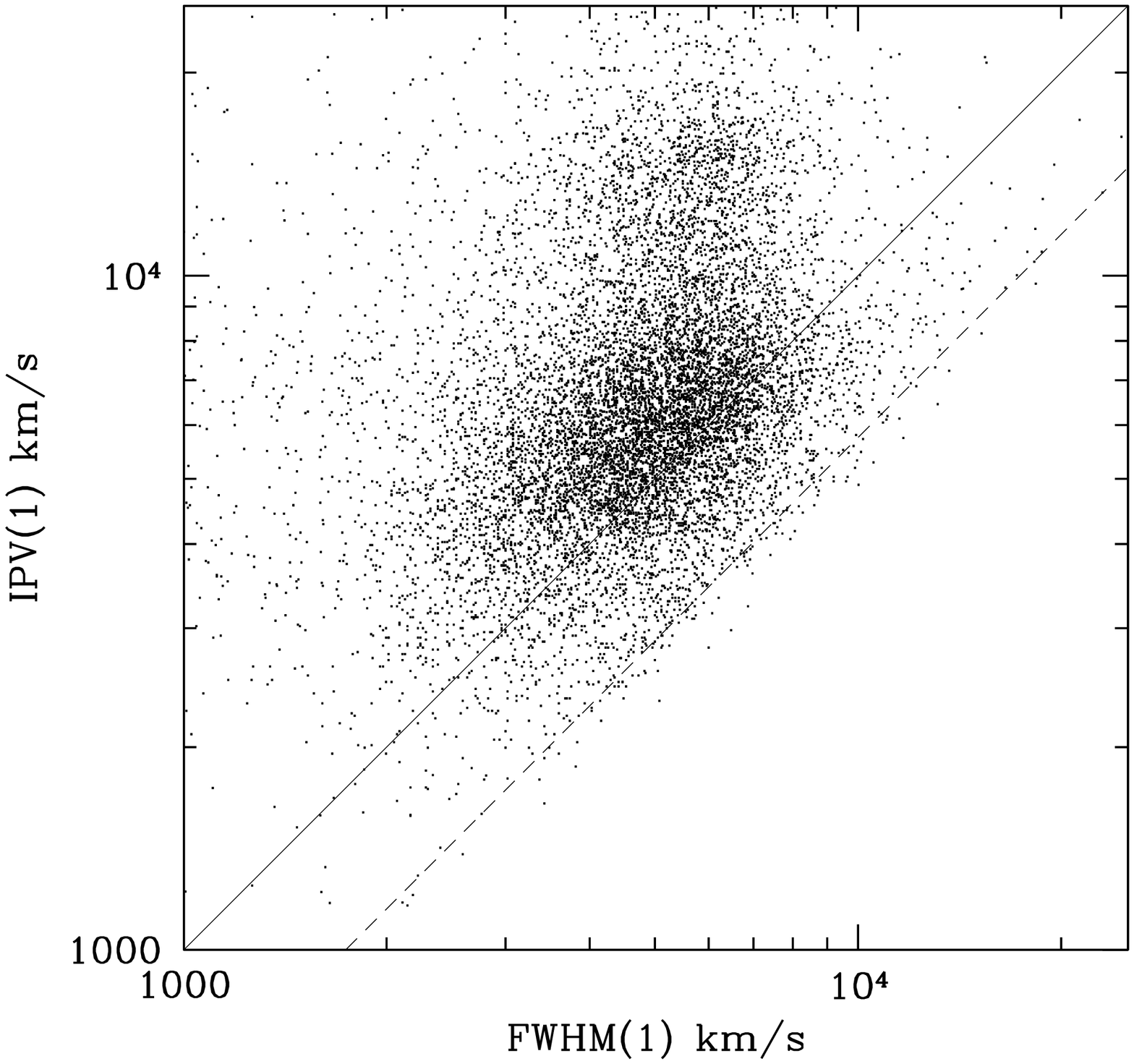,width=6.0cm}\hspace{0cm}\psfig{file=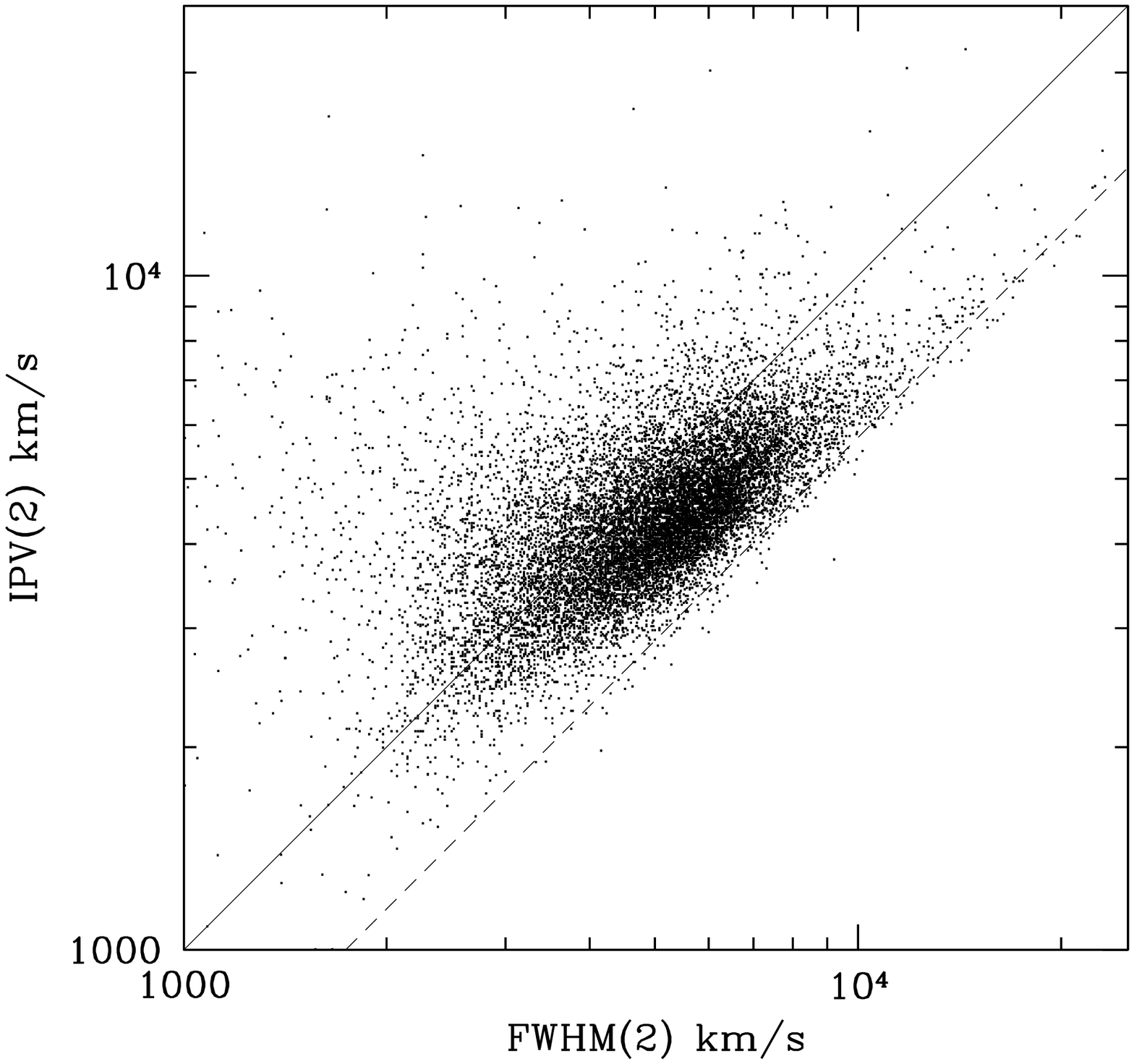,width=6.0cm}\hspace{0cm}\psfig{file=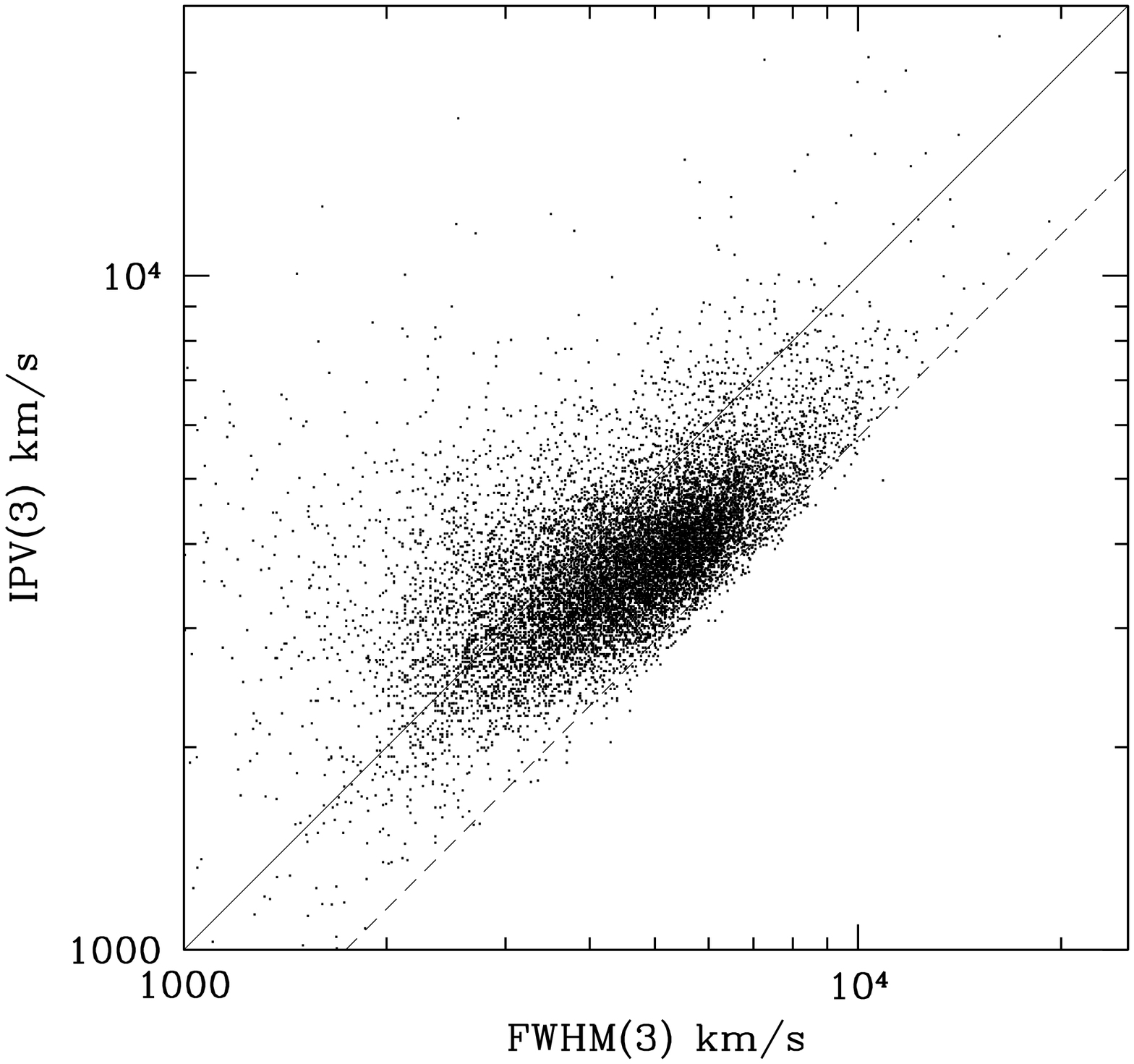,width=6.0cm}}
\caption{Comparisons between IPV width and FWHM for each of the
objects in our sample when applying the three
fitting proceedures discussed in the text. The solid line gives the 1:1
relation while the dashed line shows the ratio $\rm FWHM/IPV=1.75$
appropriate for a Gaussian.}
\label{fig_comp_ip_fw}
\end{figure*}

Fig.~\ref{fig_comp_wds} suggests that FWHMs offer a robust measure of
line width that is relatively independent of the fitting technique
applied. IPV widths are strongly influenced if one takes the emission
on the red wing of \civ\ to be a part of the line itself; however, they are
relatively robust with respect to the fitting proceedures if not. This
leaves a question as to how well FWHM and IPV width measurements
correlate with each other. Fig.~\ref{fig_comp_ip_fw} compares these
measurements for each of the fitting proceedures.

%
%

The dashed lines in Fig.~\ref{fig_comp_ip_fw} show the ratio
$\rm FWHM/IPV=1.75$ that is applicable for a Gaussian line. This line
represents a hard limit, and a two-Gaussian model cannot have 
$\rm FWHM/IPV>1.75$. The fact that we do
see scatter over the line is due to the IPV width being measured
directly from the spectrum, while the FWHM is measured from the model
fit.


Fig.~\ref{fig_comp_ip_fw}\empha\ shows that the FWHM and IPV widths do
not correlate when using fitting proceedure (1). This prescription leaves the
\civ\ line with a strong wing which affects the IPV width more than
the FWHM producing a strong skew towards larger IPV widths at any
FWHM.

Fig.~\ref{fig_comp_ip_fw}\emphb\ and \emphc\ differ from \empha. In
these plots the IPV widths and FWHMs correlate well with $75-80$\,\% of
the points lying between the lines at $\rm FWHM/IPV=1$ and 1.75. However,
both \emphb\ and \emphc\ show a significant number of points with low
FWHMs in comparison with their IPV widths.

Visual inspection of the spectra of objects with large FWHM/IPV ratios
reveals that the outlying points
represent a mix of objects. There are a small number of BAL objects in
this area 
which have been missed by our automated BAL rejection process (see
appendix~\ref{sec:bal}). In addition, there are a number low S/N
spectra in which the double Gaussian fit results in a narrow Gaussian
being fitted
to a noise spike in the spectrum which severely narrows the FWHM. Lastly,
there are a small number of objects which genuinely show very peaky
profiles with a broad underlying emission and so have a small FWHM/IPV
ratio.

\begin{figure*}
\centering
\centerline{\psfig{file=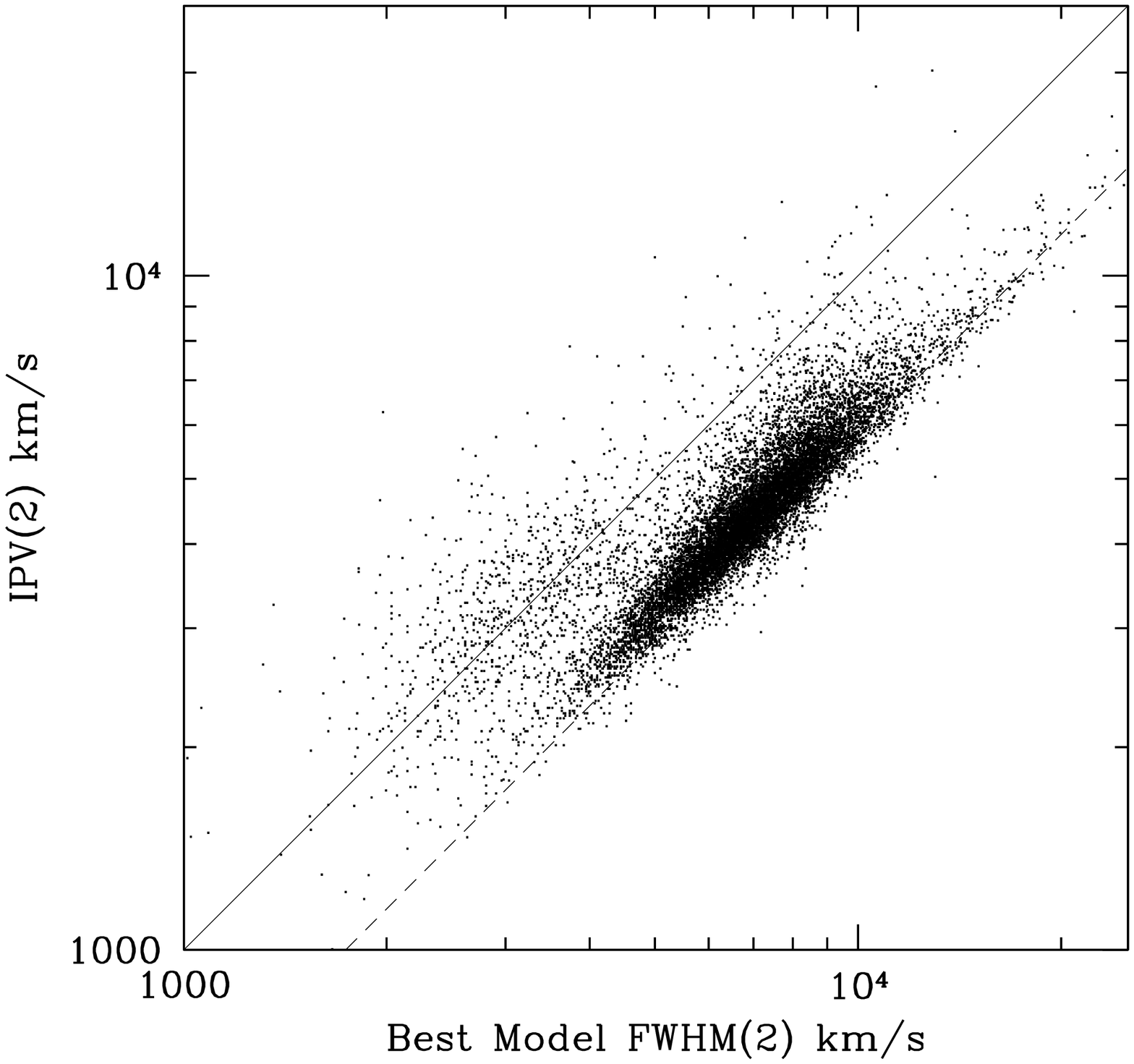,width=6.0cm}\hspace{0.5cm}\psfig{file=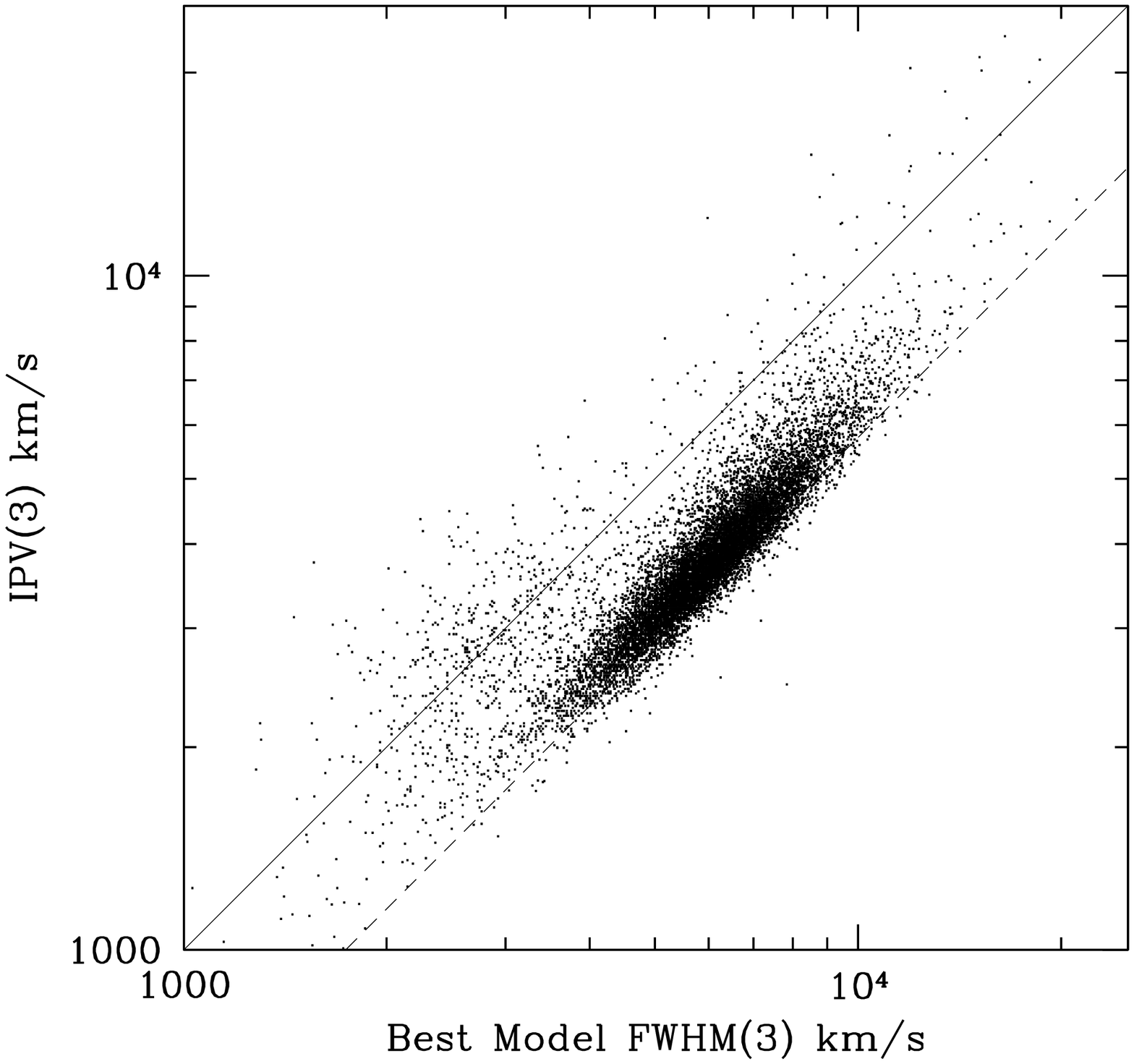,width=6.0cm}}
\caption{Comparisons between IPV widths and best model FWHMs for
fitting proceedures (2) and (3). The solid line gives the 1:1
relation while the dashed line shows the ratio $\rm FWHM/IPV=1.75$
appropriate for a Gaussian.}
\label{fig_comp_ip_bmfw}
\end{figure*}

In light of the number of low S/N objects which have FWHMs affected
by fits to noise spikes, it seems we are allowing too many degrees
of freedom in our fitting to low S/N objects. In fitting proceedures
(2) and (3), where the \civ\ line is symmetric, we also perform a
single Gaussian fit to each line. We calculate the reduced $\chi^2$
for each of these fits, then take as the best model FWHM that of the
single Gaussian unless the double Gaussian fit improves the reduced
$\chi^2$ by more than one. Fig.~\ref{fig_comp_ip_bmfw} compares the
IPV width with this best model FWHM.
Here we can see that the IPV width and best model FWHM correlate
strongly. There are a small number of outliers with small FWHM/IPV ratios,
most of which represent truly peaky \civ\ line profiles.

\subsection{Summary of line fitting techniques}

\begin{figure}
\centering
\centerline{\psfig{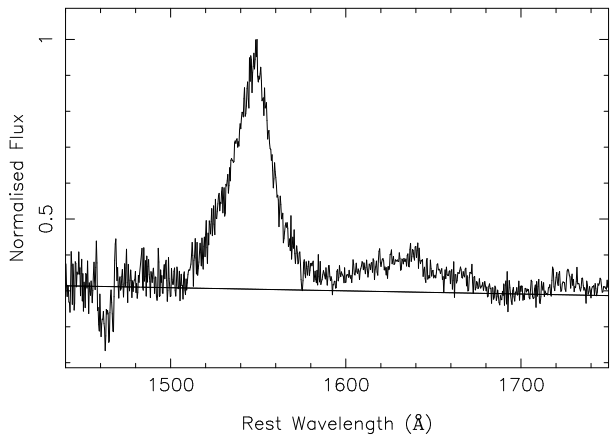}}
\caption{\civ\ region of the SDSS spectrum of object
J024634.09-082536.1. A continuum has been fitted between $\sim$1500 and
1700\,\AA. The emission redwards of
$\sim$1600\,\AA\ can hardly be considered to be associated with the
\civ\ line.}
\label{fig_civ_wing_eg}
\end{figure}

Of the three line fitting techniques we have outlined, we prefer
prescription (2). Prescription (1) results in a large red wing on
the \civ\ line and we find no evidence that this emission is genuinely
associated with \civ. Indeed there are examples of spectra which
show a dip in emission between the \civ\ line and a bulge of
emission redwards of 1600\,\AA. Fig.~\ref{fig_civ_wing_eg} shows the
spectrum of QSO J024634.09-082536.1 taken as part of the SDSS. The dip
in emission to approximately the continuum
level at 1600\,\AA\ makes it difficult to associate the emission
redwards of 1600\,\AA\ with the \civ\ line. 
It appears that the FWHM as measured via precription (1) gives results
similar to that from the other fitting proceedures. However, the IPV
widths are strongly affected by the red wing of the line.

Prescription (3) has the advantage of simplicity. However, it is overly
simplistic to the extent it can produce a systematic bias to our
results. Furthermore, fitting a linear continuum between predetermined
points either
side of the \civ\ line limits any measurements made on the line to be
within these limits, and very broad lines could be affected. While
there are very few lines broad as this we believe prescription (2) serves
better to describe both the \civ\ emission, and the emission
surrounding the line.

In the analysis that follows we use fitting proceedure (2).

\subsection{Testing the line fitting routine}

Fitting proceedure (2) is relatively complex and we need to be confident
in our results, in particular for low S/N spectra.
To test the effect of S/N on the accuracy of our fitting routine we
take the highest S/N spectra from our SDSS and 2dF samples. We add
random Gaussian noise to these spectra and then re-measure the \civ\
line width in the degraded spectrum. For each original spectrum we
add six different levels of noise, and repeat the measurement 100
times using a different random seed.

\begin{figure*}
\centering
\centerline{\psfig{file=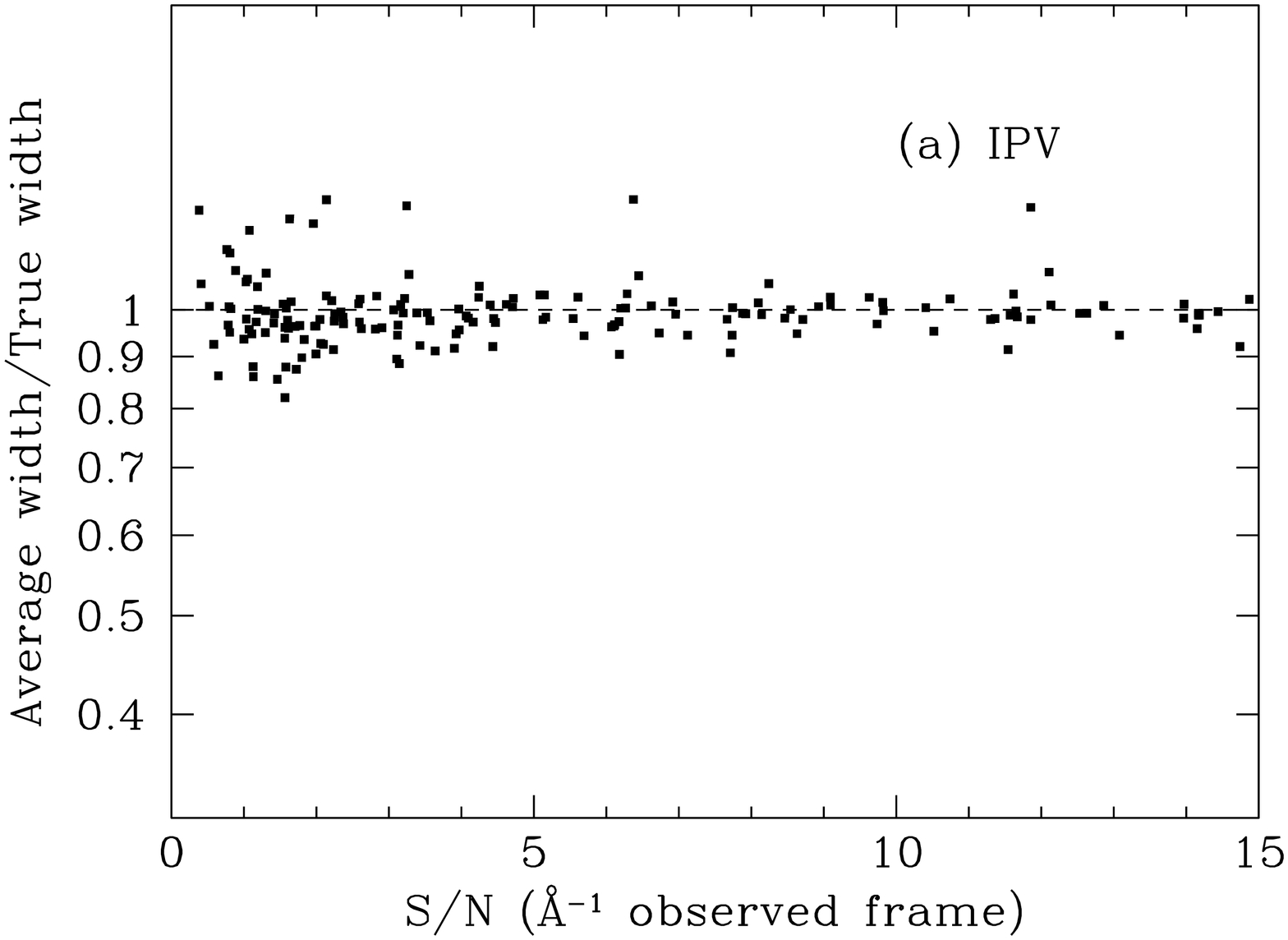,width=7.0cm,angle=-90}\psfig{file=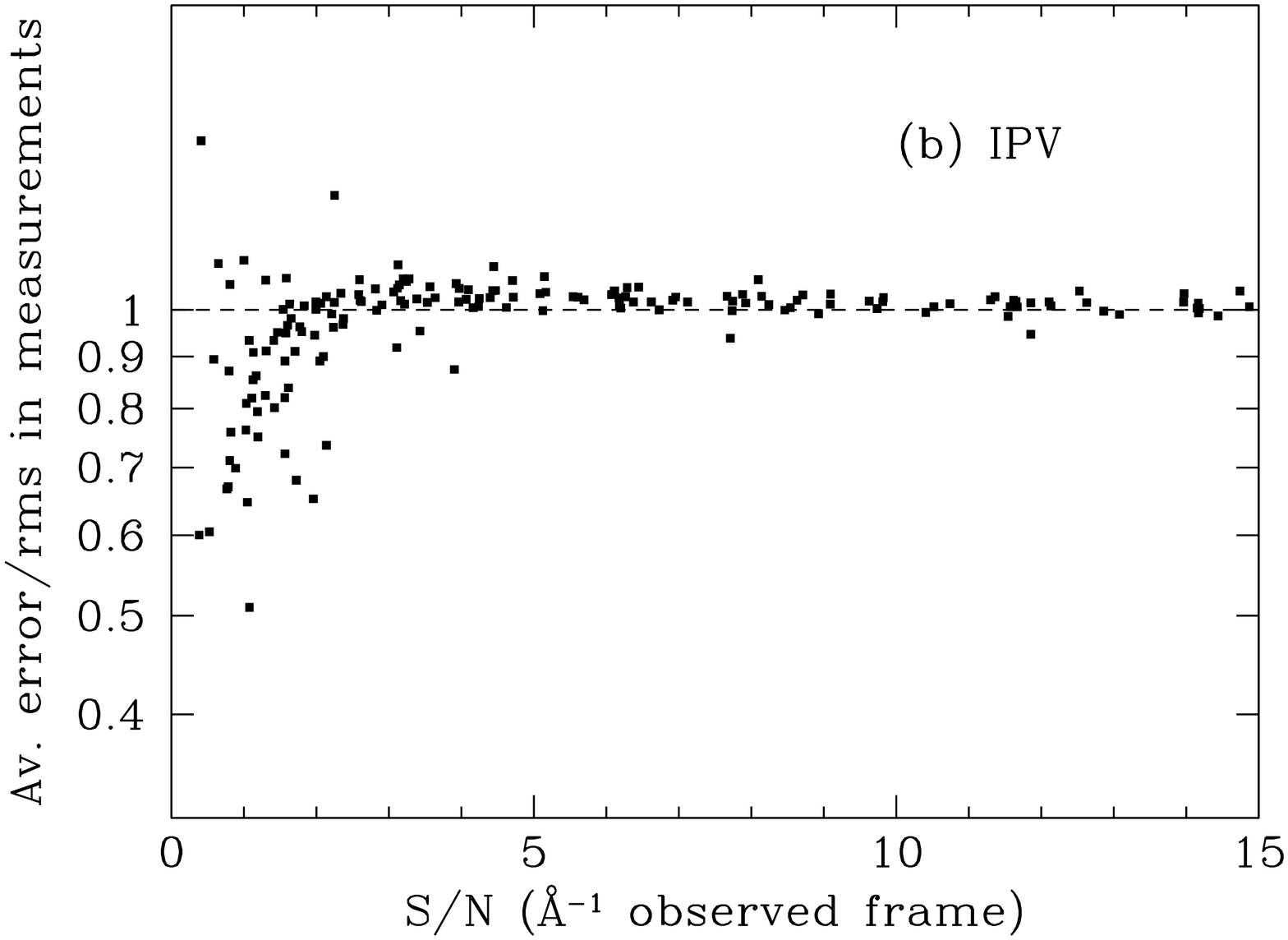,width=7.0cm,angle=-90}}
\centerline{\psfig{file=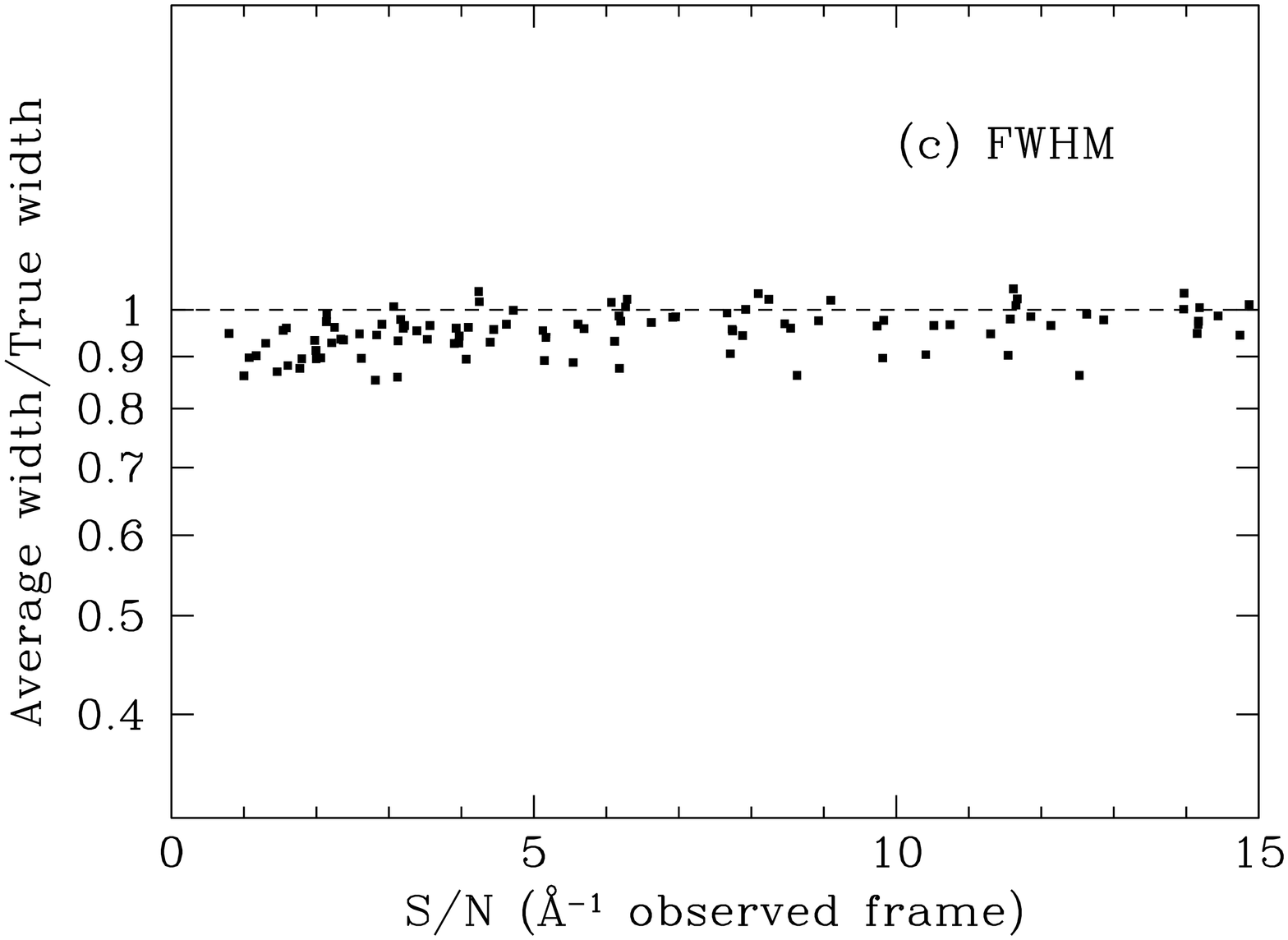,width=7.0cm,angle=-90}\psfig{file=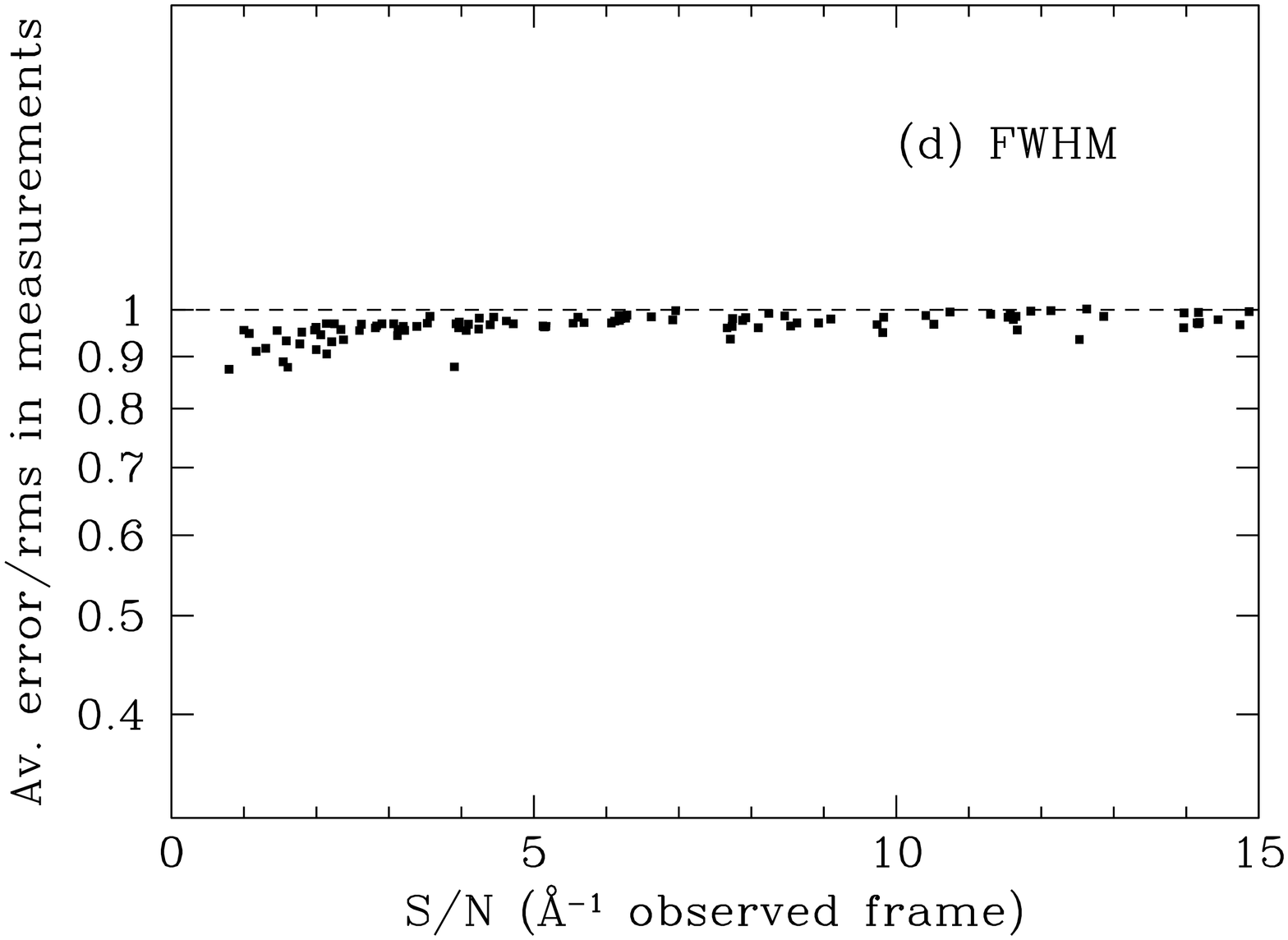,width=7.0cm,angle=-90}}
\caption{\empha\ and \emphc\ compare the average line width measured
from 100 noisy spectra with the width measured from the
original high S/N SDSS spectrum. \emphb\ and \emphd\ compare the average
error on these measurements with the rms of the line widths. \empha\
and \emphb\ compare IPV width measurements and \emphc\ and
\emphd\ compare the best model FWHM. In each plot we give the results
as a function of S/N.}
\label{fig:noise_test_c_s}
\end{figure*}

In Fig.~\ref{fig:noise_test_c_s} \empha\ and \emphc\ we compare the
average line width measured in the 100 degraded spectra with the line
width measured from the original spectrum for the IPV and best model
FWHM respectively. In \emphb\ and \emphd\ we compare the rms in the
100 line width measurements with the average error on the
measurements. Fig.~\ref{fig:noise_test_c_s} only shows results for
high S/N SDSS spectra, results for 2dF spectra show almost identical
results.

Down to a $\sn\sim3$\,\AA\pmo\ both measures are relatively stable
with respect to S/N. In general we find that (for $\sn>3$\,\AA\pmo) the
IPV width provides a less biased line width measurement, and more
accurate errors ($<$5\,\% offset in each at $\sn=3$\,\AA\pmo) when
compared to the best model FWHM. In the analysis in this paper we
will take the IPV width as the primary line width measure, and apply a
S/N cut of $\sn>3$\,\AA\pmo\ that rejects $\sim10$\,\% of our sample.





\section{BAL rejection} \label{sec:bal}

Broad absorption features alter the appearance of emission lines and
make accurate measurements of the profile impossible. The \civ\ line
is more commonly affected by BAL systems than lower
ionisation lines (e.g. \mgii; \citealt{tru06}), and when fitting \civ\
one must carefully detect and reject these QSOs from the analysis.

\subsection{Balnicity and Absorption indexes}

The traditional method for measuring the strength of broad absorption
in quasar spectra is with the balnicity index (BI; \citealt{wey91}). More
recently \citet{tru06} made a catalogue of BAL objects in the SDSS DR3
using the slightly different absorption index (AI). Both of these indicators
identify broad absorption troughs by comparing with fitted template
spectra. Regions in a spectrum in which consecutive pixels fall below 90\,\%\
of the fitted template, over a continuous region exceeding some range
in velocity, are considered broad absorption lines. The range
varies: 2000 and 1000\,km/s for the BI and AI respectively, as do the
regions in which the search for these absorption troughs is carried
out: $3000-25000$\,km/s for the BI and $0-29000$\,km/s for the
AI (for both indexes the BAL search is only carried out bluewards of
\civ). Once the broad absorption troughs around a line have been
identified, the indexes themselves are essentially the cumulative
equivalent width of these troughs.

From their definition it is clear that the BI is a more strict definition
of whether an object has broad absorption. The AI
detects narrower absorbers and searches for them over a wider velocity
range. The differences between these two indexes are discussed in
\citet{tru06} who found that, for the \civ\ line in SDSS DR3 quasars,
10\,\%\ had a non-zero BI compared with 26\,\%\ with an AI.

We find that both the BI and AI are too strict when
identifying BAL objects and we have developed our own
system for rejecting absorption systems. Our process
does not equate to a new method for identifying `definite' BALs
(see e.g. \citealt{kni08}), but is an automated method
for identifying spectral lines with absorption features
which could affect the profile of the line. Indeed many
obviously narrow-line absorbers are also rejected with our
technique.

\subsection{BAL identification}

We use two procedures for identifying BALs in our data. The first is
similar to the BI or AI, the second uses pixel binning to search for
troughs in the spectra.

\subsubsection{Method 1: Consecutive pixels}

We do not fit quasar templates to our spectra as in the BI or AI
processes. However, we do fit Gaussian models to the lines as part of
the fitting procedure. We use the best Gaussian model in a manner similar to
the templates in the BI/AI methods. We search for pixels which lie more than
$1\sigma$ below the model within $\rm \pm2FWHM$ of the fit to the
line. Any spectrum which has consecutive pixels
below this value spanning $>750$\,km/s is discarded as a potential BAL
system.

The two main differences between our method and the BI/AI methods
are the use of $1\sigma$ as the limit to define a `low'
pixel, and our use of 750\,km/s as the width threshold for defining broad
absorption. We loosen our definition of broad absorption to 750\,km/s
simply to be sure of rejecting any spectra which are significantly
affected by absorption. A deep absorption
trough 750\,km/s wide can still seriously affect line profile
measurements and, while these may not represent true BAL systems, they
are contaminants to our data.

We use $1\sigma$ rather than a fixed percentage of the
model flux because it has a simpler statistical
interpretation. A velocity of 750\,km/s represents $\sim10$ pixels in
SDSS spectra 
and $3-4$ pixels in a 2dF spectrum. In a 2dF spectrum this criterion
represents a $>99$\,\% confidence level for detecting consecutive
pixels which are genuinely deviant from the model fit.

This BAL identification routine is effective at finding obvious BAL
objects, and BAL
systems in high S/N spectra. However, our dataset contains many
objects which exhibit lower level broad absorption, often at low S/N,
which are missed by the routine. We miss BALs for two reasons. Firstly,
our model fit to the emission line is affected by the absorption
trough, making it less likely to find consecutive pixels below the
model fit.
Second, the necessity for a large
number of consecutive pixels to lie below the limit is a
very strict constraint. In objects which show only low level absorption,
it is likely that one or more of the several pixels affected by
the absorption will be scattered to within $1\sigma$ of the model fit
by random noise.

One can relax the criterion for identifying pixels affected by
absorption. However, if we loosen the criterion too
far we begin to reject high S/N objects which have systematic
residuals when compared with the model fit to the emission line.

We require a second method for identifying these low level, low
S/N BAL systems.

\subsubsection{Method 2: Binning the spectrum}

When inspecting a large number of spectra with low level broad
absorption it became clear
that the eye is capable of finding BALs when an automated
routine struggles for two reasons: firstly the eye can automatically
smooth a spectrum which reduces noise, and secondly it is sensitive to
the steep sides of an absorption trough.

We have constructed a second BAL
identification routine that relies on rebinning the spectrum onto a
larger pixel scale (i.e. constructing spectra with a larger dispersion
in terms of \AA/pix).
The rebinned spectrum has a higher S/N per pixel than the
original and, by comparing a pixel with those on either side of it in
the rebinned spectrum, we can identify whether it has been affected by
absorption.

We rebin the spectrum onto a number of pixel scales between 650 and
1000\,km/s ($\sim10-15$\,pixels in SDSS spectra, $\sim3-5$\,pixels
2dF spectra). For each pixel in the rebinned spectrum we interpolate
between the two pixels either side. If the central pixel falls below
the interpolated value by more than a given amount we reject the object as
a BAL quasar. After some experimentation we find that $5.5\sigma$ makes
for a good cutoff in SDSS spectra. The lower S/N and dispersion of 2dF
spectra means that the cutoff level must be reduced to $3.5\sigma$ to be
sure of detecting BALs.

\begin{figure*}
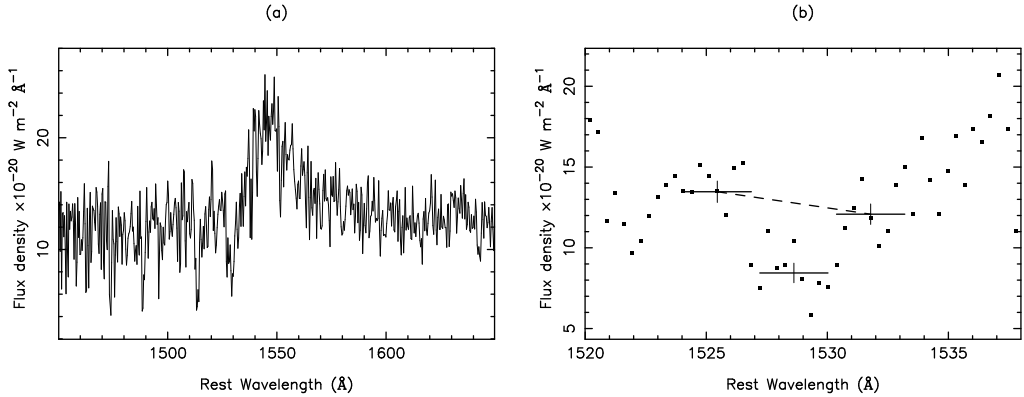
 
\centering
\centerline{\psfig{file=bin_BAL_1_1.ps,width=6.5cm,angle=-90}\hspace{0.5cm}\psfig{file=bin_BAL_1_2.ps,width=6.5cm,angle=-90}}
\caption{\empha\ The \civ\ region of the spectrum of SDSS object
J011229.41+151213.9. In \emphb\ we expand the region of the spectrum around
       1530\,\AA. The points show the original spectrum. The crosses
show three points in the rebinned spectrum. The width of each cross
       indicates the width
of the bin and the height shows the error on the rebinned flux. The
dashed line is interpolated between the pixels on either side of the
central pixel.}
\label{fig_bin_bal1}
\end{figure*}

We illustrate our second BAL identification scheme in
Fig.~\ref{fig_bin_bal1}. Fig.~\ref{fig_bin_bal1}\empha\ shows the
\civ\ region of the SDSS spectrum of object J011229.41+151213.9. In
Fig.~\ref{fig_bin_bal1}\emphb\ the absorption trough at 1530\,\AA\ is
expanded. The points in the plot are the original SDSS spectrum. The crosses
show three points from the rebinned spectrum. The width of each cross
indicates the bin size and the height shows the error on the rebinned
flux density. The dashed line is interpolated between the pixels on either
side of the central pixel which deviates by more than $5.5\sigma$ from
the line.

\begin{figure*}
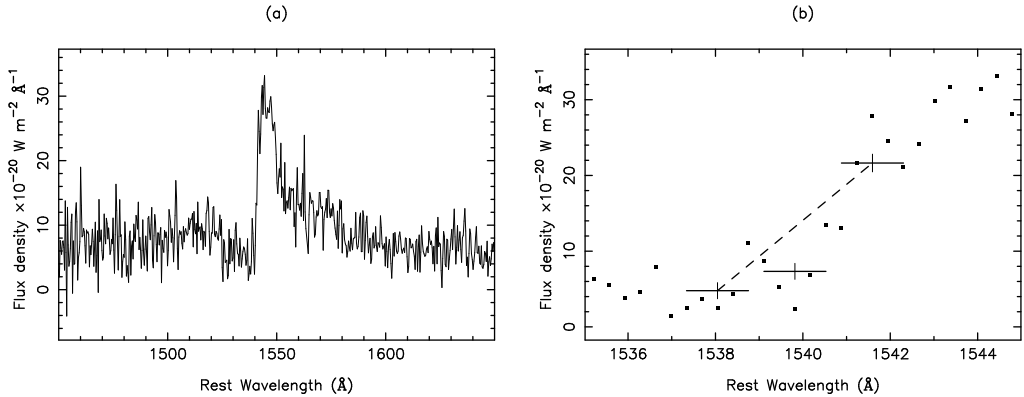
 
\centering
\centerline{\psfig{file=bin_BAL_2_1.ps,width=6.5cm,angle=-90}\hspace{0.5cm}\psfig{file=bin_BAL_2_2.ps,width=6.5cm,angle=-90}}
\caption{SDSS spectrum of quasar J010810.52+001755.8 together with an
expanded plot of the BAL region and the local rebinned spectrum. Symbols in
plots are as in Fig.~\ref{fig_bin_bal1}.}
\label{fig_bin_bal2}
\end{figure*}

The BAL identification scheme not only works for objects such as
J011229.41+151213.9, in which the absorption trough dips below the
unabsorbed spectrum on either side of the trough,
but also in BAL spectra where the absorption only drops
below the unabsorbed spectrum on one side.
Fig.~\ref{fig_bin_bal2}\empha\ shows
the spectrum of J010810.52+001755.8. The blue wing of the \civ\ line is
heavily absorbed. However, the absorption is such
that the flux density does not dip significantly below the continuum
level on the blue side of the absorption trough.
Our method of rebinning the spectra to search for BALs will also
detect the absorption in this spectrum by detecting the sharp increase
in flux density around 1540\,\AA.

In fact, since most BALs are more than 1000\,km/s wide, this method of
identifying BALs most commonly works by finding their steep edges. We
do not rebin the spectra onto scales greater than 1000\,km/s because
as we approach the width of the emission lines we find that many
non-BAL objects are also rejected.
To stop false identifications of non-BAL objects
the gradient of the rebinned spectrum must be slowly
varying across the emission line. As the bin width
approaches the width of the emission line it no longer
becomes smooth, and we will begin to identify non-BAL objects.

High S/N spectra, in particular those with narrower emission lines, can
be wrongly identified as BAL objects with this routine. While
the number of objects falsely identified as having BALs is small, the
resulting bias will be towards broader line, lower S/N
objects. However, these spectra can be identified and returned to the
sample.

During the fitting procedure we fit both single and double
Gaussian profiles to the emission line. Quasar emission lines are not
well modelled by single Gaussians and, in the case of high S/N spectra,
they have large reduced $\chi^2$s. The double Gaussian models, however,
fit the emission lines relatively well and have reduced $\chi^2$s
around unity. We find that the vast majority of spectra which are
wrongly classified as having BALs by our routine have a significantly
improved fit to the double Gaussian model for the emission line
compared to the single Gaussian fit (i.e. the difference between the
reduced $\chi^2$s is greater than one).

\vspace{0.2cm}
Our procedure for rejecting BALs from our sample is then as
follows. We perform both of our BAL detection routines (both searching
for consecutive low pixels and the binning technique). For the objects
which are rejected as having BALs we compare the reduced $\chi^2$s for single
and double Gaussian fits to the emission line. All those which have a
significant improvement in the fit when using two Gaussian to model
the lines are then visually inspected and any non-BAL objects are
returned to the sample.

\subsection{Testing the BAL rejection code}

To gain an impression of how well our routine works at removing BAL
objects from our sample, as well as how many non-BAL objects may be
rejected, we have visually inspected 1,000 spectra
from the SDSS and identified them as BAL or non-BAL
objects.

Of the 1000 objects, 356 were visually identified as having BALs, of these the
automated routine rejects 326. In total
358 objects were rejected by the BAL identification routine with 32
non-BAL objects rejected because they have high S/N narrow
lines. In 27 of these false-positive identifications the \civ\ line is
significantly better
fit with a double Gaussian model compared to a single Gaussian and so
would be returned to our sample after visual inspection.

After performing our BAL rejection procedure on these
1,000 objects, 669 are passed  by the procedure as not having
BALs. Included in these 669 objects are 30 unidentified BAL systems
(4\,\%), and we have rejected 5 non-BAL objects
in a biased manner ($<1$\,\%).

It is worth pointing out that our BAL identification
process also rejects a significant number of narrow absorption
objects. In some of these spectra the line profile is relatively
unaffected by the absorption, and line properties could be derived
from the spectrum. However, since these are rejected in an unbiased
manner in terms of the emission line properties they are counted as
acceptable BAL rejections.

\end{document}